\documentclass[usenatbib,fleqn]{mnras}

\makeatletter
\newlength{\abovecaptionskip}%
\makeatother

\usepackage{mathtools}
\usepackage{threeparttable}

\usepackage{amsmath,amssymb}
\usepackage{cases}
\usepackage{mathrsfs}
\usepackage[]{subfig}
\usepackage{float} 
\usepackage{graphicx}
\usepackage{epstopdf}
\graphicspath{{./figures/}}
\usepackage{url}

\newcommand\lsim{\mathrel{\rlap{\lower4pt\hbox{\hskip1pt$\sim$}}
    \raise1pt\hbox{$<$}}}
\newcommand\gsim{\mathrel{\rlap{\lower4pt\hbox{\hskip1pt$\sim$}}
    \raise1pt\hbox{$>$}}}

\newcommand       \be          {\begin{eqnarray}}
\newcommand       \ee          {\end{eqnarray}}
\newcommand{\Mbh}[1][]{M_{\bullet#1}}
\newcommand{\Menc}{M_{\rm enc}}

\newcommand{\Msun}{{\rm M_\odot}}

\newcommand{\rs}{r_s}

\begin{document}
\title[Influence of CNM on TDE radio emission]{The influence of
  circumnuclear environment on the radio emission from TDE jets}
\author[Generozov et al.]{
  A. Generozov$^{1}$\thanks{ag@astro.columbia.edu},
  P. Mimica$^{2}$, B. D. Metzger$^{1}$,
  N. C. Stone$^{1}$,
  D. Giannios$^{3}$, M.~A. Aloy$^{2}$
  \\
  $^{1}$Columbia Astrophysics Laboratory, Columbia University, 550 West 120th Street, New York, NY 10027\\
  $^{2}$Departamento de Astronom\'{\i}a y Astrof\'{\i}sica, Universidad de Valencia, E-46100 Burjassot, Spain\\
  $^{3}$Department of Physics and Astronomy, Purdue University, 525
  Northwestern Avenue, West Lafayette, IN 47907, USA
}

\maketitle
\begin{abstract}
  Dozens of stellar tidal disruption events (TDEs) have been
  identified at optical, UV and X-ray wavelengths.  A small fraction
  of these, most notably {\it Swift} J1644+57, produce radio
  synchrotron emission, consistent with a powerful, relativistic jet
  shocking the surrounding circumnuclear gas.  The dearth of similar
  non-thermal radio emission in the majority of TDEs may imply that
  powerful jet formation is intrinsically rare, or that the conditions
  in galactic nuclei are typically unfavorable for producing a
  detectable signal.  Here we explore the latter possibility by
  constraining the radial profile of the gas density encountered by a
  TDE jet using a one-dimensional model for the circumnuclear medium
  which includes mass and energy input from a stellar population.
  Near the jet Sedov radius of 10$^{18}$ cm, we find gas densities in
  the range of $n_{18} \sim$ 0.1$-$1000 cm$^{-3}$ across a wide range
  of plausible star formation histories.  Using one- and
  two-dimensional relativistic hydrodynamical simulations, we
  calculate the synchrotron radio light curves of TDE jets (as viewed
  both on and off-axis) across the allowed range of density profiles.
  We find that bright radio emission would be produced across the
  plausible range of nuclear gas densities by jets as powerful as {\it
    Swift} J1644+57, and we quantify the relationship between the
  radio luminosity and jet energy.  We use existing radio detections
  and upper limits to constrain the energy distribution of TDE jets.
  Radio follow up observations several months to several years after the TDE
  candidate will strongly constrain the energetics of any relativistic
  flow.
\end{abstract}


\section{Introduction}
\label{sec:intro}
When a star in a galactic nucleus is deflected too close to the
central supermassive black hole (BH), it can be torn apart by tidal
forces.  During this tidal disruption event (TDE), roughly half of the
stellar debris remains bound to the BH, while the other half is flung
outwards and unbound from the system.  The bound material, following a
potentially complex process of debris circularization
(\citealt{Kochanek1994,Guillochon+2013,Hayasaki+2013,Hayasaki+2016,Shiokawa+2015,Bonnerot+2016}),
accretes onto the BH, creating a luminous flare lasting months to
years \citep{Hills1975, Carter+1982, Rees1988}.

Many TDE flares have now been identified at optical/ultraviolet (UV)
\citep{Gezari+2008, Gezari+2009, van-Velzen+2011, Gezari+2012,
  Arcavi+2014, Chornock+2014, Holoien+2014, Vinko+2015, Holoien+2016}
and soft X-ray wavelengths \citep{Bade+1996, Grupe+1999,
  Komossa&Greiner1999, Greiner+2000, Esquej+2007, Maksym+2010,
  Saxton+2012}. Beginning with the discovery of {\it Swift} J1644+57
(hereafter SwJ1644) in 2011, three additional TDEs have been
discovered by their hard X-ray emission (\citealt{Bloom+2011,
  Levan+2011, Burrows+2011, Zauderer+2011, Cenko+2012, Pasham+2015,
  Brown+2015}).  Unlike the optical/UV/soft X-ray flares, these events
are characterized by non-thermal emission from a transient
relativistic jet beamed along our line of sight, similar to the blazar
geometry of active galactic nuclei (AGN).  In addition to their highly
variable X-ray emission, which likely originates from the base of the
jet (see e.g. \citealt{Bloom+2011, Crumley+2016}), these events are
characterized by radio synchrotron emission \citealt{Berger+2012,
  Zauderer+2013, Cenko+2012}\footnote{Swift J1112.2 8238 was not
  promptly followed up in the radio, but subsequent follow-up with
  ATCA shows radio emission at a much higher level than expected for
  the galaxy's UV/emission line luminosities (Andrew Levan, private
  communication).}.  The latter, more slowly evolving, is powered by
shocks formed at the interface between the jet and surrounding
circumnuclear medium (CNM)
\citep{Bloom+2011,Giannios&Metzger2011,Metzger+2012,De-Colle+2012,Kumar+13,Mimica+2015},
analagous to the afterglow of a gamma-ray burst.

Although a handful of jetted TDE flares have been observed, the
apparent volumetric rate is a very small fraction ($\sim
10^{-5}-10^{-4}$) of the observed TDE flare rate (e.g.,
\citealt{Burrows+2011}, \citealt{Brown+2015}), and an even smaller
fraction of the theoretically predicted TDE rate
\citep{Wang&Merritt2004,Stone&Metzger2016}.  One explanation for this
discrepancy is that the majority of TDEs produce powerful jets, but
their hard X-ray emission is relativistically beamed into a small
angle $\theta_{\rm b}$ by the motion of the jet, making them visible
to only a small fraction of observers.  However, the inferred beaming
fraction $f_b \approx \theta_{b}^{2}/2 \sim 10^{-5}-10^{-4}$ would
require $\theta_{\rm b} \sim 0.01$ and hence a jet with a bulk Lorentz
factor of $\Gamma \gtrsim 1/\theta_{\rm b} \sim 100$, much higher than
inferred for AGN jets or by modeling SwJ1644
(\citealt{Metzger+2012}). This scenario 
would also require an unphysically low jet half opening angle
$\theta_j\lsim 0.01$.

The low detection rate of hard X-ray TDEs may instead indicate that powerful
jet production is intrinsically rare, or that the conditions in
the surrounding environment are unfavorable for producing bright
emission.  Jets could be rare if they require, for instance, a highly
super-Eddington accretion rate (\citealt{De-Colle+2012}), a TDE from a
deeply plunging stellar orbit (\citealt{Metzger&Stone2016}), a TDE in
a retrograde and equatorial orbit with respect to the spin of the
black hole \citep{Parfrey+2015}, or a particularly strong magnetic
flux threading the star (\citealt{Tchekhovskoy+2014,Kelley+2014}).
Alternatively, jet formation or its X-ray emission could be suppressed
if the disk undergoes Lens-Thirring precession due to a misalignment
between the angular momentum of the BH and that of the disrupted star
(\citealt{Stone&Loeb2012}).  In the latter case, however, even a
`dirty' jet could still be generated, which would produce luminous
radio emission from CNM interaction.

\citet{Bower+2013} and \citet{van-Velzen+2013} performed radio
follow-up of optical/UV and soft X-ray TDE flares on timescales of
months to decades after the outburst (see also
\citealt{Arcavi+2014}). They detect no radio afterglows definitively
associated with the host galaxy of a convincing TDE
candidate.\footnote{There were radio detections for two ROSAT flares:
  RX J1420.4+5334 and IC 3599. However, for RX J1420.4+5334 the radio
  emission was observed in a different galaxy than was originally
  associated with the flare.  IC 3599 has shown multiple outbursts in
  the recent years, calling into question whether it is a true TDE at
  all \citep{Campana+2015}. The optical transient CSS100217 (see
  \citealt{Drake+2011}) had a weak radio afterglow, but its peak
  luminosity is more consistent with a superluminous supernova than a
  TDE.} \citet{Bower+2013} and \citet{van-Velzen+2013} use a Sedov
blast wave model for the late-time radio emission to conclude that
$\lesssim 10\%$ of TDEs produce jetted emission at a level similar to
that in SwJ1644.  \citet{Mimica+2015} use two-dimensional
(axisymmetric) hydrodynamical simulations, coupled with synchrotron
radiation transport, to model the radio emission from SwJ1644 as a jet
viewed on-axis.  By extending the same calculation to off-axis viewing
angles, they showed that, regardless of viewing angle, the majority of
thermal TDE flares should have been detected if their jets were as
powerful as SwJ1644, which had a total energy of $\sim 5\times
10^{53}$ erg.

The recent TDE flare ASSASN-14li (\citealt{Holoien+2016a}) was
accompanied by transient radio emission, consistent with either a weak
relativistic jet \citep{van-Velzen+2016} or a sub-relativistic outflow
\citep{Alexander+2016,Krolik+2016} of total energy $\sim
10^{48}-10^{49}$ erg.  The 90 Mpc distance of ASSASN-14li, a few
times closer than most previous TDE flares, implies that even if other
TDEs were accompanied by similar emission, their radio afterglows
would fall below existing upper limits.  The extreme contrast between
the radio emission of SwJ1644 and ASSASN-14li indicates that the
energy distribution of TDE jets is very broad.

Previous works (\citealt{Bower+2013}; \citealt{van-Velzen+2013};
\citealt{Mimica+2015}) have generally assumed that all TDE jets
encounter a similar gaseous environment as SwJ1644.  However, the
density of the circumnuclear medium (CNM) depends sensitively on the
input of mass from stellar winds and the processes responsible for
heating the gas (\citealt{Quataert2004,Generozov+2015}). 

The first goal of this paper is to constrain the range of gas
densities encountered by jetted TDEs using the semi-analytic model for
the CNM ($\S\ref{sec:cnm}$) developed in \citet{Generozov+2015}
(hereafter GSM15).  With this information in hand, in
$\S\ref{sec:results}$ we present hydrodynamical simulations of the
jet-CNM shock interaction which determine the radio synchrotron
emission across the allowed range of gaseous environments, for
different jet energies and viewing angles.  In $\S\ref{sec:param}$ we
show how the dependence of our results for the peak luminosity, and
time to radio maximum, on the jet energy and CNM density can be
reasonably understood using a simple analytic blast wave model
($\S\ref{sec:analyt}$, Appendix~\ref{app:analyt}), calibrated to the
simulation data.  Then, using extant radio detections and upper
limits, we systemtically constrain the energy distribution of TDE
jets.  One of our primary conclusions is that TDE jets as energetic as
SwJ1644 are intrinsically rare, a result with important implications
for the physics of jet launching in TDEs and other accretion flows.
Our work also lays the groundwork for collecting and employing future,
larger samples of TDEs with radio follow-up, to better constrain the
shape of the energy distribution.  We summarize and conclude in
$\S\ref{sec:conc}$.

\section{Diversity of CNM Densities}
\label{sec:cnm}

\subsection{Analytic Constraints}
\label{sec:analy}

Jet radio emission is primarily sensitive to the density of ambient
gas near the Sedov radius, $r_{\rm sed}$, outside of which the jet has
swept up a gaseous mass exceeding its own. For a power law gas density
profile, $n= n_{18} \left(r/10^{18} {\rm cm}\right)^{-k}$,
\begin{align}
  r_{\rm sed} &= 10^{18} \,{\rm cm}\, \left( \frac{E(3-k)}{4\pi n_{18}
      m_{\rm p} c^2 (10^{18}\,{\rm cm})^3} \right)^{1/(3-k)}  \nonumber\\
  &\approx 3 E_{54}^{1/2} n_{\rm 18}^{-1/2}\,{\rm pc}.
  \label{eq:rdec}
\end{align}
where $E = E_{54}10^{54}$ erg is the isotropic equivalent energy and
in the final equality we have taken $k = 1$, typical of our results
described later in this section.  For a powerful jet similar to
SwJ1644, the deceleration radius is typically of order a parsec, but
it can be as small as $10^{16}$ cm for a weak jet/outflow, such as
that in ASASSN-14li.

Although an initially relativistic jet will slow to sub-relativistic
speeds at $r \sim r_{\rm sed}$, significant deceleration already sets
in at the deceleration radius (where the jet has swept up a fraction
$\sim 1/\Gamma$\footnote{This is really the Lorentz factor of the
  shock (see \citealt{Hascoet+2014}). For simplicity, we use the
  Lorentz factor of the ejecta, which leads to a factor of $\sim
  2$ underestimate of the deceleration time.} of its rest mass),
\begin{equation}
  r_{\rm dec}=\frac{r_{\rm sed}}{\Gamma^{2/(3-k)}}.
  \label{eq:rdec2}
\end{equation}
According to an observer within the opening angle of the jet, the jet
reaches the Sedov and deceleration radii, respectively, at times given
by
\begin{equation}
t_{\rm sed} \simeq \frac{r_{\rm sed}}{c} \approx
10 E_{54}^{1/2}n_{18}^{-1/2} {\rm year}
 \end{equation} 
\begin{equation}
t_{\rm dec} \simeq
\frac{r_{\rm dec}}{2\Gamma^{2} c} = \frac{t_{\rm
    sed}}{2\Gamma^{2(4-k)/(3-k)}} = \frac{t_{\rm sed}}{2\Gamma^{3}},
 \label{eq:tdec}
\end{equation}
where in the final equality we have again taken $k = 1$.

\subsubsection{Dynamical Model of CNM}
\label{sec:model}

In the absence of large scale inflows, the dominant source of gas in
the CNM of quiescent galaxies is winds from stars in the galactic
nucleus. We bracket the range of possible nuclear gas densities using
a simple steady-state, spherically symmetric, hydrodynamic model
including mass and energy injection from stellar winds. The relevant
equations are (e.g. \citealt{Holzer+1970}; \citealt{Quataert2004})
\begin{align}
  &\frac{\partial \rho}{\partial t}+\frac{1}{r^2}\frac{\partial}{\partial r}\left(\rho r^2 v\right)=q \label{eq:drhodt}\\
  &\rho \left(\frac{\partial v}{\partial t} + v\frac{\partial
      v}{\partial r}\right) =-\frac{\partial p}{\partial r}- \rho\frac{GM_{\rm enc}}{r^{2}} -q v \label{eq:dvdt}\\
  &\rho T\left(\frac{\partial s}{\partial t} + v\frac{\partial
      s}{\partial
      r}\right)=q\left[\frac{v^2}{2}+\frac{\tilde{v}_w^2}{2}-\frac{\gamma_{\rm
      ad}}{\gamma_{\rm ad}-1}
    \frac{p}{\rho} \right] ,
\label{eq:model}
\end{align}
where $\rho $, $v$, $p$, and $s$ are the density, velocity, pressure
(we assume an ideal gas with a mean molecular weight of $0.62$ and
adiabatic index $\gamma_{\rm ad}$=5/3), and specific entropy of the
gas, respectively. The enclosed mass $\Menc = M_{\bullet} + M_{\star}$
includes both the black hole mass $M_{\bullet}$ and enclosed stellar
mass $M_{\star} \propto \int \rho_{\star}r^{2}dr$, where
$\rho_{\star}$ is the stellar density.  At the radius of the sphere of
influence, $r_{\rm inf}$, the enclosed stellar and black masses are
equal, $M_{\star}(r_{\rm inf})=\Mbh$.  We take $r_{\rm inf}=3.5
\Mbh[,7]^{0.6}$ pc (GSM15), where $\Mbh[,7]=\Mbh/10^7 \Msun$.

The source term $q$ is the mass injection rate per unit volume per
unit time. We take $q=\eta \rho_{\star}/t_h$, where $\eta$ is a
dimensionless efficiency parameter that depends on the properties of
the stellar population and $t_h$ is the Hubble time. The
$\tilde{v}_w^2=\sigma(r)^2+v_w^2$ term in the entropy equation is the
specific heating rate of the gas per unit volume, where 

\begin{equation}
\sigma \approx \sqrt{\frac{3 G \Mbh}{(\Gamma+2)
    r}+\sigma_\star^2},
\label{eq:sigmarel}
\end{equation}
is the stellar velocity dispersion, which approaches the constant
value of $\sigma_\star$ outside of the influence radius. As in GSM15
we have taken $\sigma_\star=190 \Mbh[,7]^{0.2} {\rm km s^{-1}}$ (based
on the $\Mbh-\sigma$ relation from
\citealt{McConnell+2011})\footnote{This may be of questionable
  validity for low mass black holes (e.g. \citealt{Greene+2010,
    Kormendy&Ho2013}). Also, several of the black hole masses used in
  \citet{McConnell+2011} were underestimated
  \citep{Kormendy&Ho2013}. However, the precise form of the
  $\Mbh-\sigma$ relationship has minimal impact on our
  results}. $v_w^2$ is the specific heating rate of the gas from other
sources including stellar wind kinetic energy, supernovae, and black
hole feedback. We take $v_w$ to be independent of radius.

GSM15 present analytic approximations for the
densities and temperatures of steady state solutions to
equation~\eqref{eq:model}. We apply these results across the
physically allowed range of heating ($v_w$) and mass injection
rates ($\eta$), and obtain the corresponding range of gas densities.

\subsubsection{Stellar density profiles}
We assume a broken power law for the stellar density profile,
$\rho_{\star}$, motivated by Hubble measurements of the radial surface
brightness profiles for hundreds of nearby early type galaxies
\citep{Lauer+2007}.  The measured profile is well fit by the so-called
``Nuker'' law parameterization, i.e.~a piece-wise power law that smoothly
transitions from an inner power law slope, $\gamma$, to an outer power
law slope, $\beta$, at a break radius, $r_b$.

Most galaxies have $0<\gamma<1$, and are classified into two broad
categories: ``core'' galaxies with $\gamma<0.3$ and ``cusp'' galaxies with
$\gamma>0.5$. Assuming spherical symmetry and a constant mass-to-light
ratio, the inner stellar profile translates to a stellar density of
$\rho_\star\propto r^{-1-\gamma}=r^{-\delta}$. 

Cusp-like stellar density profiles are the most relevant to TDEs,
since as described in \citet{Stone&Metzger2016}, a cuspy stellar
density profile results in a higher TDE rate per galaxy.  We adopt a
fiducial value of $\gamma=0.7$ ($\delta=1.7$), motivated by the
rate-weighted average value of the inner stellar density profile for
the galaxies in \citet{Stone&Metzger2016} (their Table C).

\subsubsection{Gas density profiles}
\label{sec:dProf}
\begin{figure}
\includegraphics[width=8cm]{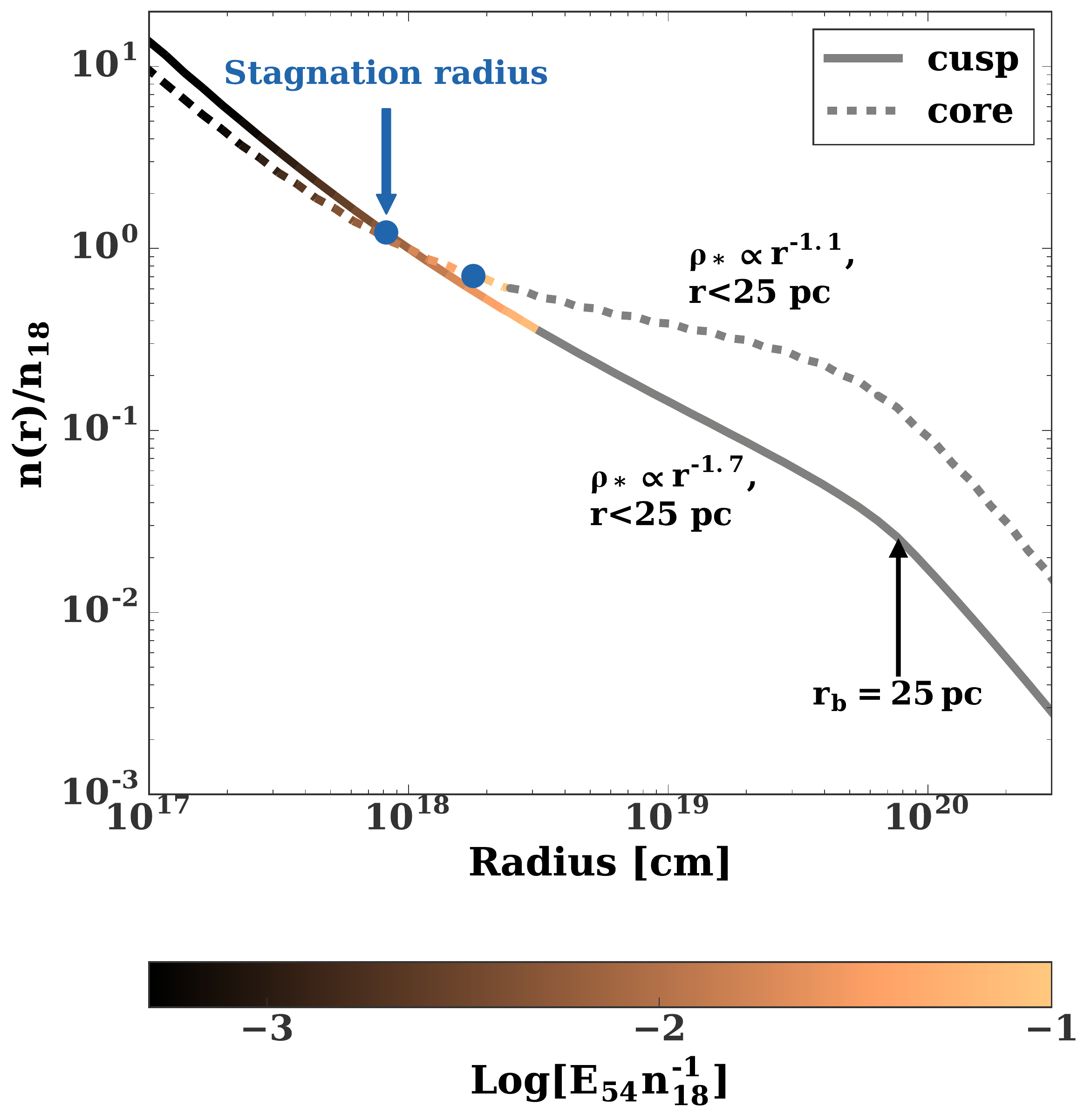}
\caption{\label{fig:profiles} Steady-state radial profiles of the CNM
  gas density, normalized to its value at $10^{18} {\rm cm}$,
  $n_{18}$. The profiles are calculated for a black hole mass of
  $10^{7} \,\Msun$ and a gas heating parameter of $v_w=600$ km
  s$^{-1}$.  Cusp and core stellar density profiles are shown with
  solid and dashed lines, respectively.  The line colors denote the
  ratio of isotropic equivalent jet energy to $n_{18}$ which results
  in $r = r_{\rm sed}$ at each radius.}
\end{figure}

Given sufficiently strong heating, a one-dimensional steady-state
model for the CNM is characterized by an inflow-outflow structure.
The velocity passes through zero at the ``stagnation radius'', $\rs$.
Mass loss from stars interior to the stagnation radius flows inwards,
while that outside of $\rs$ is unbound in an outflow from the nucleus.
Fig.~\ref{fig:profiles} shows example radial profiles of the
steady-state gas density calculated for a core and a cusp stellar
density profile. The stagnation radius is marked as a blue dot on each
profile.

As long as the heating parameter, $v_w$, is greater than the
stellar velocity dispersion,
\begin{align}
r_s \simeq f(\delta) \frac{G M_{\bullet}}{v_w^2}
  \simeq 0.4 \Mbh[,7] v_{500}^{-2} \,{\rm pc},
\label{eq:rs}
\end{align}
where $v_{500}\equiv v_w/500 \,{\rm km\, s^{-1}}$ and $f(\delta)$ is a
constant of order unity, which in the second equality we take equal to
its fiducial value of $f(\delta = 1.7)$=2.5 (see GSM15). The gas density at the
stagnation radius, $n(\rs)$, is determined by the rate at which
stellar winds inject mass interior to it,
\begin{equation}
  \dot{M}=\frac{\eta M_{\rm \star}(\rs)}{t_h} \approx  2.8 \times 10^{-6} \Mbh[,7]^{0.22} \eta_{0.02} \left(\frac{r_s}{\rm
      pc}\right)^{1.3} \Msun \, {\rm yr}^{-1},
\label{eq:dotM}
\end{equation}
where $M_{\star}(\rs)$ is the total stellar mass enclosed within the
stagnation radius, $\eta_{0.02}=\eta/0.02$ is normalized to a value
characteristic of an old stellar population, and the second equality
again assumes our fiducial value of $\delta=1.7$.

The density at the stagnation radius, $n(\rs)$, is estimated by
equating the gas injected by stellar winds over a dynamical time at
the stagnation radius, $t_{\rm dyn} (\rs)$, to
the gas mass enclosed at this location.  
\begin{align}
  &\frac{4 \pi}{3} \rs^3  m_p n(r_s) \simeq \dot{M} t_{\rm dyn}
  (\rs)
\end{align}
For $\rs<r_{\rm inf}$, $t_{\rm dyn}=(\rs^3/G \Mbh)^{1/2}$, while for
$\rs>r_{\rm inf}$, $t_{\rm dyn}=(\rs/\sigma_\star)$. Thus,
\begin{align}
  & n(r_s) \simeq
\begin{cases}
0.1 \eta_{0.02} \Mbh[,7]^{-0.28}
  \left(\frac{r_s}{\rm pc}\right)^{-0.2} {\rm cm}^{-3} & \rs<r_{\rm
    inf}\\
0.1 \eta_{0.02} \Mbh[,7]^{0.02} \left(\frac{r_s}{\rm pc}\right)^{-0.7}
{\rm cm}^{-3} & \rs>r_{\rm
    inf},
\end{cases}
\label{eq:nrs}
\end{align}
For sufficiently strong heating, the stagnation radius will lie inside
the SMBH's sphere of influence and will be given by
equation~\eqref{eq:rs}. In this case,
\begin{equation}
n(r_s) \simeq 0.2 \, v_{500}^{0.4} \eta_{0.02} \Mbh[,7]^{-0.48} {\rm cm}^{-3},
\label{eq:nrs2}
\end{equation}
Near the stagnation radius, GSM15 found that the radial gas profile
has a power-law slope of $k \approx (4\delta-1)/6$, which for our
fiducial value of $\delta=1.7$ gives $n \propto r^{-1}$. The gas
density steepens towards smaller radii, approaching $n\propto
r^{-1.5}$, for radii well inside of both the stagnation radius of the
flow and the SMBH's sphere of influence. The gas profile flattens to
$n\propto r^{1-\delta}$ between the stagnation radius and the stellar
break radius; however, for our fiducial value of $\delta = 1.7$, the
resulting profile $n\propto r^{1-\delta} \approx r^{-0.7}$ is only
moderately changed. We expect at the deceleration radius of most jets
is bracketed by $r^{-0.7}$ and $r^{-1.5}$. For simplicity we adopt
\begin{equation} n(r)= n_{18} \left(\frac{r}{10^{18}
    {\rm cm}}\right)^{-1},
\label{eq:profile}
\end{equation}
as our fiducial density profile, where $n_{18}$ is the density at $r =
10^{18}$ cm. We explore the effects of the density slope on jet radio
emission in $\S$~\ref{sec:profileComp}

\citet{Alexander+2016} use radio observations of the ASSASN-14li flare
to infer a nuclear gas density profile of $n\propto r^{-2.6}$ for its
host galaxy on scales of $\sim 10^{16}$ cm--much steeper than
our fiducial density profile. However, we note that this galaxy was
active before the flare, possibly explaining the unusually steep
density profile.

Combining equations \eqref{eq:nrs} and (\ref{eq:profile}),
we obtain
\begin{equation}
  n_{18}\simeq
\begin{cases}
 0.4 \left(\frac{r_s}{\rm pc}\right)^{0.8}
  \Mbh[,7]^{-0.28} \eta_{0.02} \, {\rm cm^{-3}} & r_s<r_{\rm inf}\\
 0.4 \left(\frac{r_s}{\rm pc}\right)^{0.3}
  \Mbh[,7]^{0.02} \eta_{0.02} \, {\rm cm^{-3}} & r_s>r_{\rm inf}.
\end{cases}
  \label{eq:n18}
\end{equation}
For sufficiently strong heating, the stagnation radius will lie inside
the sphere of influence and will be given by
equation~\eqref{eq:rs}. In this case,
\begin{equation}
  n_{18}\simeq 0.2 \Mbh[,7]^{0.52} v_{500}^{-1.6} \eta_{0.02} \, {\rm
    cm^{-3}}.
\label{eq:n182}
\end{equation} 
As shown in Fig.~\ref{fig:profiles}, the gas density profile 
steepens outside the break radius $r_b$ of the stellar density
profile.  However, this will only impact the radio emission near its
maximum if $r_b$ lies inside of the Sedov radius, $r_{\rm sed}$
(eq.~\ref{eq:rdec}).  The lines in Fig.~\ref{fig:profiles} are colored
according to the combination of jet energy and CNM density $n_{18}$
which results in $r = r_{\rm sed}$ at each radius.  The measured break
radii of all but four of the \citet{Lauer+2007} galaxies exceed 10
parsecs, which greatly exceeds $r_{\rm sed}$ even in the case of a
very energetic jet ($E=4\times 10^{54}$ erg) in a low density CNM of
$n_{18} \sim 1$ cm$^{-3}$.  The presence of a nuclear star cluster
(NSC) in the galactic center could produce another break in the
stellar density profile near the outer edge of the cluster, which is
typically located at $r_{\rm nsc} \sim 1-5$ pc \citep{Georgiev+2014}.
But even in this case, only particular combinations of high $E$/low
$n_{\rm 18}$ result in $r_{\rm sed} > r_{\rm nsc}$.  We therefore
neglect the effects of an outer break in the stellar density profile
in our analysis.

\subsubsection{Allowed Density Range}
\label{sec:densAllowed}
We now estimate the allowed range in the normalization of the CNM gas
profile, $n_{18}$.  We assume that star formation occurs in two
bursts, an old burst of age comparable to the Hubble time $t_{\rm h} =
10^{10}$ yr, and a ``young'' burst of variable age $t_{\rm burst} \ll
t_{\rm h}$ which contributes a fraction $f_{\rm burst}$ of the stellar
mass. We assume a Salpeter IMF for both stellar populations.

For a sufficiently large burst of age $\lesssim$ 40 Myr, gas heating
is dominated by the energetic winds of massive
stars.\footnote{Core-collapse SNe are also an important heating
  source.  In a young stellar population, the power from core-collapse
  supernovae exceeds that from massive stellar winds after $\sim$6 Myr
  \citep{Voss+2009}. However, due to discreteness effects the heating
  from massive star winds will be more important on small scales.}  In
this case the mass return ($\eta$) and heating parameters ($v_w$) are
calculated as described in Appendix C of GSM15. Given $\eta(t_{\rm
  burst},f_{\rm burst})$ and $v_w(t_{\rm burst},f_{\rm burst})$, we
calculate $n_{18}$ following equation~\eqref{eq:n182}.

For an older stellar population, a few different sources contribute to
gas heating, including Type Ia Supernovae (SNe)\footnote{Unbound
  debris streams from TDEs potentially provide another source of
  heating localized in the galactic center
  (\citealt{Guillochon+2016}), which we neglect.} and AGN feedback.
We focus on quiescent phases, during which SNe Ia dominate.  As
discussed in GSM15, SNe Ia clear out the gas external to a critical
radius, $r_{\rm Ia}$, where the interval between successive Ia SNe
equals the dynamical (gas inflow) timescale.  For an old stellar
population, $n_{18}$ is estimated by equating $r_{\rm Ia}$ with the
stagnation radius in equation~\eqref{eq:n18}.  The
Ia radius is calculated as described in GSM15 at times $t>300 \,{\rm
  Myr}$ after star formation, and is taken to be constant for $t =
40-300$ Myr.\footnote{GSM15 incorrectly extrapolated the Ia rate valid
  at times $t>300 \,{\rm Myr}$ back to a time $t = 3$ Myr, which is
  unphysical as no white dwarfs would have formed by this time.
  Although its qualitative impact on our results is minimal, here we
  instead take the Ia rate to be 0 for $ t < 40$ Myr.}

\begin{figure} 
  \includegraphics[width=8cm]{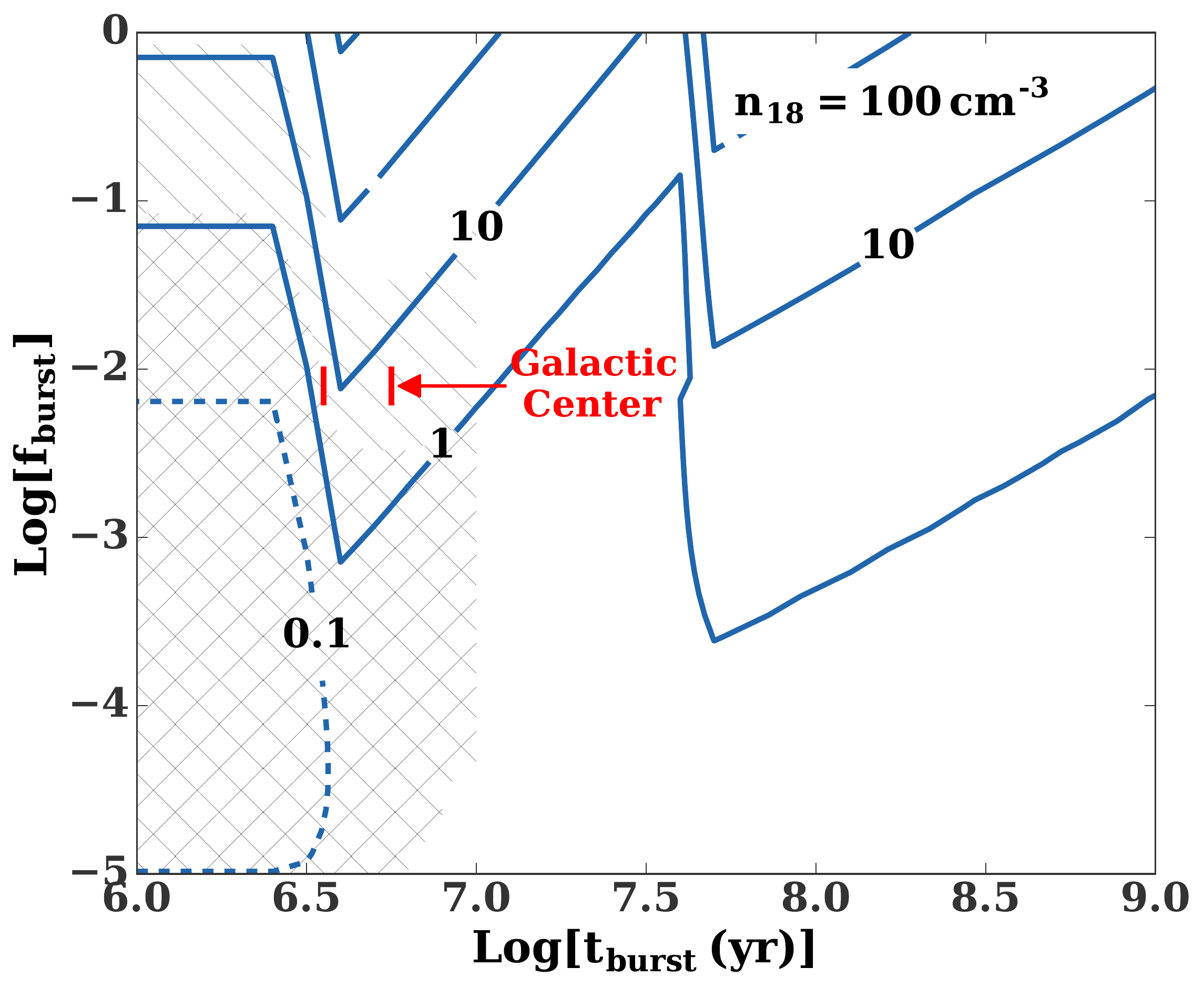}
  \caption{\label{fig:param} Contours of $n_{18}$, the CNM density at
    $r = 10^{18}$ cm (blue lines), as a function of the stellar
    population in the galactic nucleus.  The star formation is
    parameterized assuming that a fraction $f_{\rm burst}$ of the
    stars form in a burst of age $t_{\rm burst}$, while the remaining
    stars formed a Hubble time ago.  We have assumed a black hole mass
    of $10^{7} \, \Msun$ and that both the young and old stars possess
    a cusp-like density profile, with a corresponding gas density
    profile $n \propto r^{-1}$.  Hatched areas indicate regions of
    parameter space where massive stars ($\gsim 15 \, \Msun$) dominate
    the gas heating rate, but less than one (doubly hatched) or less
    than ten (singly hatched) massive stars are present on average
    inside the nominal stagnation radius (eq.~\ref{eq:rs}).  In these
    regions discreteness effects not captured by our formalism are
    potentially important. The red line shows the approximate location
    of the Galactic Center in this parameter space (see text for
    details).}
\end{figure}

Fig.~\ref{fig:param} shows how $n_{18}$ varies with the young
starburst properties, $f_{\rm burst}$ and $t_{\rm burst}$.  We find a
maximum density of $n_{18} \sim 1,300\, \Mbh[,7]^{0.5}$ cm$^{-3}$ is
achieved for a burst of age $t_{\rm burst} \sim 4$ Myr which forms
most of the stars in the nucleus ($f_{\rm burst} \sim 1$).  In this
case, both the energy and mass budgets of the CNM are dominated by
fast winds from massive stars.  Although a large gas density is
present immediately after a starburst, the density will decline with
the wind mass loss rate, approximately $\propto t^{-3}$, i.e. by an
order of magnitude within just a few Myr.

By contrast, the lowest allowed density $\sim 0.02 \Mbh[,7]^{0.5}$
cm$^{-3}$ is achieved for a relatively modest burst of young stars
$t_{\rm burst} \approx 10^{6}$ Myr, which forms a fraction $f_{\rm
  burst} = 4\times 10^{-4}$ of the total stellar mass. In this case
the young massive stars provide a high heating rate, while the mass
injection rate is comparatively low and receives contributions from
both young and old stars.

The lowest allowed $n_{18}$ may be an underestimate as we do not
include the effects of discreteness on the assumed stellar
population. In particular, we assume that stars provide a spatially
homogeneous heating source and mass source, even on small radial
scales where the number of massive stars present may be very small.
The doubly hatched region in Fig.~\ref{fig:param} denotes the region
where less than one massive star ($\gsim 15 \Msun$) is on average
present inside of the nominal stagnation radius (eq.~\ref{eq:rs}).
Discreteness effects are thus important for relatively small bursts of
star formation, including the case described above which gives the
minimum $n_{18}$.  If we instead equate the stagnation radius to the
radius enclosing a single star of mass $\gsim 15 \Msun$, we find a
larger value of $n_{18}\sim 0.3 \Mbh[,7]^{-0.4}$ cm$^{-3}$.  The true
minimum density therefore likely lies closer to $0.3 \Mbh[,7]^{-0.4}$
cm$^{-3}$. However, we caution that this is a very crude estimate, and
the low number of mass and heat sources means could there could be
considerable scatter about this value from stochastic variations
in the stellar population. Additionally, stellar angular momentum
could reduce the density (see e.g. \citealt{Cuadra+2006}).
 
Finally, \citet{French+2016} find that most optical/UV\footnote{We are
  not aware of any studies of host galaxy properties for the x-ray
  selected sample} TDEs have evidence of recent star formation. Six of
the eight galaxies in their sample are consistent with an
exponentially declining star formation history, forming $10\%$ of the
stars in the galaxy over $\sim 100-200$ Myr\footnote{While this paper
  was in press \cite{French+2016a} presented a more detailed study of
  stellar populations of TDE hosts, showing that their recent
  starbursts are older and smaller than we assume here, reducing the
  expected n$_{18}$ to $\sim$2 cm$^{-3}$. However, a larger density could
  still be possible if the starburst is centrally concentrated, as
  observed in nearby post-starburst galaxies (\citealt{Pracy+2012}).}.
In this region of parameter space corresponding to the right side of
Fig.~\ref{fig:param}, gas heating rate is dominated by SN Ia and
$n_{18}\sim 10$ cm$^{-3}$.

In summary, the CNM densities of quiescent galaxies vary from
min($n_{18}) \sim 0.3 \Mbh[,7]^{-0.4}$ cm$^{-3}$ to max($n_{18})\sim
1.3\times 10^{3} \Mbh[,7]^{0.5}$ cm$^{-3}$, with a characteristic value
of $n_{18}\sim 10$ cm$^{-3}$ expected for TDE host galaxies. 

\subsubsection{Mass drop-out from star formation?}

Our CNM model predicts the total gas density as sourced by stellar
winds, including both hot and cold phases.  For the first few Myr
after a starburst, the injected stellar wind material is hot ($T\gsim
10^{7}$ K) due to the thermalized wind kinetic energy.  At later
times, SNe Ia provide intermittent heating, but the stellar wind
material that accumulates on small radial scales between successive
SNe Ia may be much cooler, with at most the virial temperature $\sim
2\times 10^5 \Mbh[,7]^{0.4}$ K. This means the gas could condense into
cold clumps.

The propagation of jets through a medium containing clumps, clouds or
stars has been studied in the context of AGNs (e.g.,
\citealt{WangWiita+2000, ChoiWiita+2007}) and microquasars (e.g.,
\citealt{Araudo+2009,Perucho+2012}). It was found that the presence of
these obstacles has an effect on the long-term jet stability, as well
as observational signatures at high energies. However, the situation
is different in the case of either a very wide or ultra-relativistic
outflow (such as a GRB) for which the emission is expected to be
similar for a clumpy and a smooth medium with the same average density
(e.g.~\citealt{Nakar&Granot2007,van-Eerten+2009,Mimica&Giannios2011}).
In the case of SwJ1644, the inferred angular width of the jet
(especially of the slow component) is much larger than in the case of
AGNs and microquasars (see discussion in \citealt{Mimica+2015}).  In
fact, it is large enough to make the overall effect of the presence of
any inhomogeneities in the external medium minor. An analogous effect
is found in case of SN remnants sweeping a clumpy medium
\citep{Obergaulinger+2015}.  We note that we call the ``slow component
of the jet,'' may in fact be an unrelated mildly relativistic outflow.

On the other hand, a fraction of the cold gas may also condense into
stars. However, once the density of the hot phase is sufficiently
reduced, the cooling time will become much longer the dynamical time
and the gas will become thermally stable, causing the condensation
process to stop. For gas at the virial temperature of $\sim 2\times
10^5 \Mbh[,7]^{0.4}$ K, we find that thermal stability would be
achieved for $n_{18} \sim 0.6 \Mbh[,7]^{0.2}$ cm$^{-3}$ (where we have
defined thermal stability as the cooling time being ten times longer
than the dynamical time-scale \citealt{Mccourt+2012}). In fact this
estimate is conservative. If a fraction of the gas condenses into
stars, then feedback from stellar winds would suppress further
fragmentation. More realistically, the CNM density may be reduced by
less than a factor of $\sim$2 by star formation.

\subsubsection{Constraints from the Galactic Center}
\label{sec:empirical}

Due to its close proximity, it is possible to directly observe the gas
density distribution on parsec scales in the Galactic Center
(GC). \citet{Baganoff+2003} find that the hot, diffuse plasma within
10 arcseconds ($\sim 10^{18}$ cm) of Sgr A* has a root mean square
electron density of $\sim 26$ cm$^{-3}$.

In Fig.~\ref{fig:param} we show two sets of two-burst star formation
models which produce heating and mass return parameters comparable to
those derived from the full star formation history of the GC from
\citet{Pfuhl+2011} (their Fig.~14).  Our formalism gives values of
$n_{18}\sim 3-5$ cm$^{-3}$, too low compared to observations. 
Discrepency at this level is not surprising because our model is
spherically symmetric, while in reality many of the massive stars in
the GC are concentrated in two counter-rotating disks
\citep{Genzel+2003} with a possibly top heavy IMF \citep{Bartko+2010,
  Lu+2013}.  The disk stars extend from $\sim 10^{17}-10^{18}$ cm and
inject $\sim 10^{-3} \Msun$ yr$^{-1}$ of stellar wind material, much
more than the $\sim 4 \times 10^{-5} \Msun$ yr$^{-1}$ expected for the
global star formation history, explaining the large density of hot
gas.

In short, accurate modeling of the gas distribution in a particular
galactic nucleus, requires detailed knowledge of the distribution of
stars. Our goal here has been to bracket the range of possible nuclear gas
densities, by considering a broad range of stellar populations.

The Galactic Center also contains a cold circumnuclear ring
(e.g. \citealt{Becklin+1982}) with an opening angle of
$\sim$12$\pm3^{\circ}$ \citep{Lau+2013} and a spatially averaged
density of $\sim 10^{5}$ cm$^{-3}$ (although this varies by a few
orders of magnitude throughout the ring--see \citealt{Ferriere2012} and
references therein). Additionally, the volume from $\sim$0.4-2.5 pc
is filled with warm, ionized atomic gas with density of $\sim 900$
cm$^{-3}$ \citep{Ferriere2012}. This gas cannot be accounted for in
our model, and may originate from larger scale inflows or a disrupted
giant molecular cloud.

\begin{table}
\begin{threeparttable}
  \caption{\label{tab:jetParams} Parameters for on-axis jet simulations.}
  \begin{tabular*}{0.95\columnwidth}{lll}
\hline
& Fiducial value & Other values \\
\hline\hline
    Fast component ($\Gamma=10$) &  &  \\ 
    \hline
    $[\theta_{\rm min}$, $\theta_{\rm max}]$ & [0, 0.1] radians & \\
    $E_{\rm ISO}/10^{54}$ erg & 4  & 0.04, 0.4\\
    $E/10^{54}$ erg & 0.02 & \\
    \hline 
    Slow component ($\Gamma=2$)\\
\hline
    $[\theta_{\rm min}$, $\theta_{\rm max}]$ & [0.1, $\pi/2$] radians
    & \\
    $E_{\rm ISO}/10^{54}$ erg & $4.7$ & 0.047, 0.47 \\
    $E/10^{54}$  erg & $0.47$ & \\
    \hline
    Microphysical parameters\\
\hline
    $\epsilon_e$ & 0.1 & \\
    $\epsilon_b$ & 0.002 & \\
    $p$ & 2.3\\
    \hline 
    Nuclear gas density \\
\hline
    $n_{18}$/cm$^{-3}$ & 60 & 2, 11, 345, 2000
  \end{tabular*}
\end{threeparttable}
\end{table}

\section{Synchrotron Radio Emission}

\label{sec:results}
\subsection{Numerical Set-Up}
\label{sec:numerical}
We calculate the synchrotron radio emission from the jet-CNM shock
interaction across the physically plausible range of nuclear gas
densities.  We perform both one- and two-dimensional (axisymmetric)
relativistic hydrodynamical simulations using the numerical code
MRGENESIS \citep{MimicaGianniosAloy2009}. MRGENESIS periodically
outputs snapshots with the state of the fluid in its numerical
grid. These snapshots are then used as an input to the radiative
transfer code SPEV \citep{Mimica+2009}. SPEV detects the forward shock
at the jet-CNM interface, accelerates non-thermal electrons behind
the shock front, evolves the electron energy and spatial distribution
in time, and computes the non-thermal emission taking into account the
synchrotron self-absorption (interested readers can find many more
technical details in \citealt{Mimica+2016}). We use the same
numerical grid resolution as in \citet{Mimica+2015}.

\begin{figure}

\includegraphics[width=8.5cm]{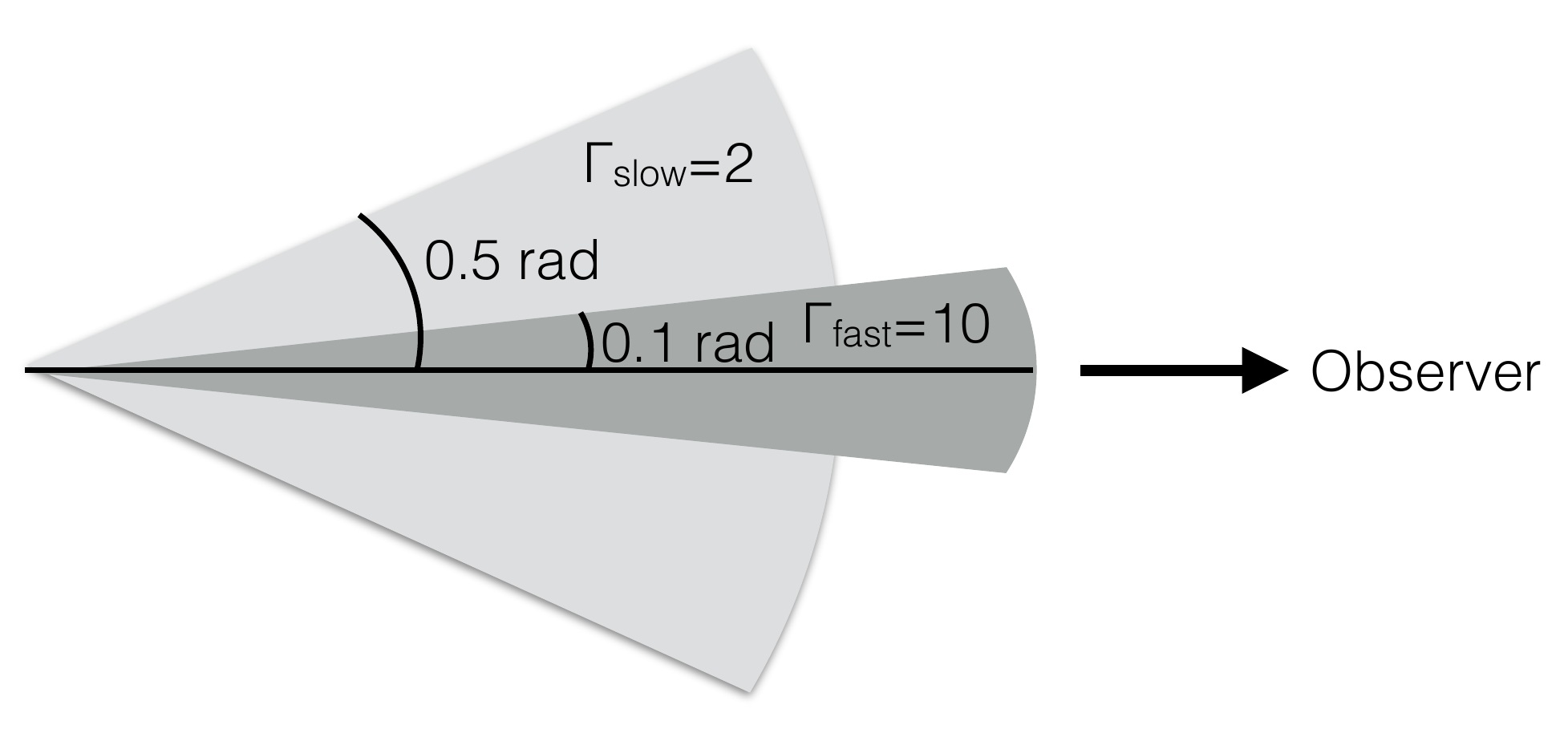}
\caption{\label{fig:jetstruct} Initial geometry of the jet used for
  our hydrodynamic simulations. We note that for 1D- two component jet
  models, we perform separate models for the inner fast core and for
  the outer sheath, which are later combined to provide the resulting
  emission. For our 1D simulation we take a slow component
  extending from 0-$\pi/2$ radians to account for the effects of jet
  spreading.}
\end{figure}

For the jet angular structure, we adopt the preferred two-component
model for SwJ1644 from \citet{Mimica+2015}, corresponding to a fast,
inner core with Lorentz factor $\Gamma = 10$, surrounded by a slower,
$\Gamma=2$ outer sheath. The ratio of the beaming-corrected energy of
the fast component is fixed to be 4\% of that of the slow sheath. A
schematic depiction of the jet geometry is shown in
Fig.~\ref{fig:jetstruct}.  In our 2D simulations the fast inner core
spans an angular interval $0-0.1\ {\rm radians}$, while the slow outer
sheath extends from $0.1\ {\rm radians}$ to $0.5\ {\rm rad}$.  The
time dependence of the jet kinetic luminosity is given by
(\citealt{Mimica+2015})
\begin{equation}\label{eq:lum}
L_{\rm j, ISO}(t) = L_{j,0}\max\left[1, (t/t_0)\right]^{-5/3},
\end{equation}
where $t_0 = 5\times 10^5$ s is the duration of peak jet power. This
is assumed to match that of the period of the most luminous X-ray
emission of SwJ1644.  Integrating equation~(\ref{eq:lum}) from $t = 0$
to $\infty$ gives the isotropic equivalent energy of the jet, $E_{\rm
  ISO}$, where $L_{j,0}=0.4\, E_{\rm ISO}/t_0$. 
For the microphysical parameters characterizing the fraction of the
post-shock thermal energy placed into relativistic electrons
($\epsilon_e$) and magnetic field ($\epsilon_B$), and the power-law
slope of the electron energy distribution $p$, we adopt the values
from the best fit model in \citet{Mimica+2015} (see
Table~\ref{tab:jetParams}).

For our 1D simulations, we modify the geometry of the slow sheath to
better mimic the results of the 2D simulations.  In our 2D models the
sheath is injected within a relatively narrow angular interval;
however, at late stages of evolution the bow shock created by the
jet-CNM interaction spans a much larger angular range due to lateral
spreading.  To account for the slow component becoming more isotropic
near peak emission in our 2D simulations \citep[bottom two panels of
Fig.~8 in][]{Mimica+2015}, we instead take the slow component to
extend from 0.1 to $\pi/2$ radians in our 1D models.  We keep the true
energy of the slow component fixed so that the isotropic equivalent
energy of the slow component is a factor of
$[\cos(0.1)-\cos(0.5))/(\cos(0.1)-\cos(\pi/2))]\approx 0.12$ smaller
than in the corresponding 2D simulations.

\begin{figure}
\includegraphics[width=8cm]{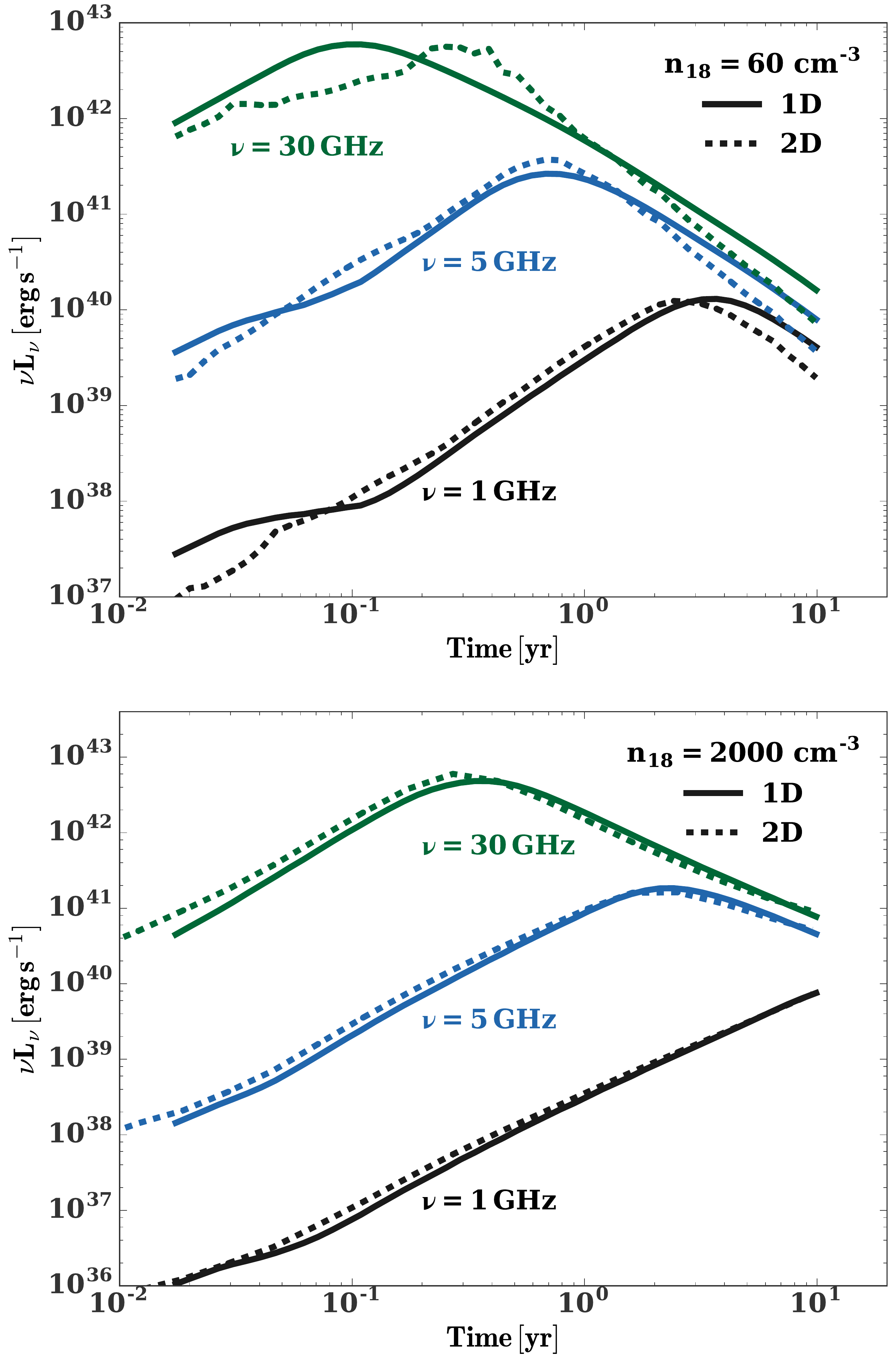}
\caption{\label{fig:1D2DB} Comparison of light curves from 1D and 2D
  simulations for an on-axis observer ($\theta_{j} = 0$). We
  assume that the gas density $n\propto r^{-1}$.}
\end{figure}

Figure \ref{fig:1D2DB} compares light curves calculated from this
modified 1D approach to the results of the full 2D simulations.
Despite the slow sheath being initially much broader in the 1D
simulations than in 2D, the resulting light curves agree surprisingly
well.  The agreement is particularly good at the highest densities
($n_{18}=2000$ cm$^{-3}$) because the slow component rapidly
isotropizes in 2D.  At lower densities ($n_{18}=60$ cm$^{-3}$), the
agreement with the 1D simulations is not as good, particularly at 30
GHz. At high densities, the jet is quickly isotropized and its
morphology is closer to that of the wedge we assume in our 1D
model. Hence, the late time evolution of the light curve at high CNM
densities is well captured by the 1D model. At lower densities,
the optically thin emission shows a strongly perturbed axially
symmetric jet, with an intricate morphology
(\citealt{Mimica+2015}). Thus, the 1D model is not optimal for
capturing the slope of the light curve, especially at the highest
frequencies (since the ejecta becomes optically thin
earlier). However, the 1D model reproduces the peak luminosity from
the 2D results within a factor of $\sim$2 for $n_{18}$=60 cm$^{-3}$
across all frequencies.

\subsection{Analytic Estimates}
\label{sec:analyt}
\begin{figure}
\includegraphics[width=8cm]{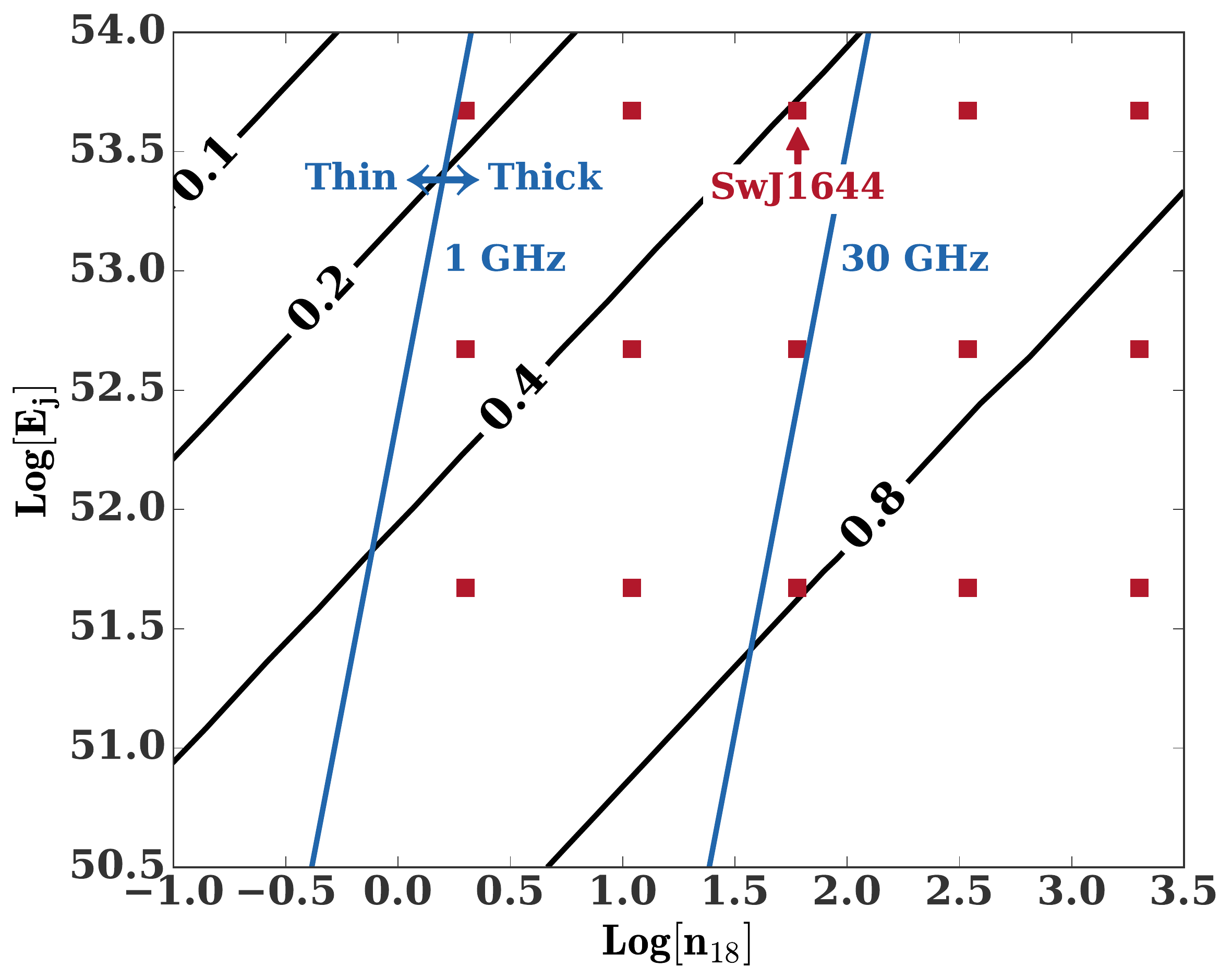}
\caption{\label{fig:diss} Contours of the fraction of the kinetic
  energy of the slow component of the jet ($\Gamma=2$) which is
  dissipated at the reverse shock in the parameter space of jet
  energy, $E_{\rm j}$, and CNM density, $n_{18}$.  The parameters of
  the suite of jet simulations presented in this paper are shown as
  red squares. The approximate location of SwJ1644 in the parameter
  space is also labeled.  Blue lines delineate the parameter space
  where the slow component of the jet is optically thin/thick at the
  deceleration time at 1 GHz (left line) and 30 GHz (right line).}
\end{figure}

The dependence of the synchrotron peak luminosity, peak time, and late
time luminosity power law slope on the ambient gas density and jet
parameters can be estimated analytically using a simple model for the
emission from a homogenous, shocked slab of gas behind a
self-similarly expanding blast wave (e.g., \citealt{Sari+98,
  Granot+02}).  The relevant results, as presented by
\citet{Leventis+2012}, are summarized in Appendix~\ref{app:analyt}.
The peak luminosity of the slow component of the jet can be estimated
from equation~(\ref{eq:peakLumGen}),
\begin{align}
&\nu L_{\nu, p}=\nonumber\\ &\text{min}
\begin{dcases}
  2.7\times 10^{40} \left(\frac{E}{10^{54} {\rm ergs}}\right)^{0.59}
  \left(\frac{\epsilon_e}{0.1}\right)^{1.3}\times \\
  \left(\frac{\epsilon_b}{0.002}\right)^{0.825}\left(\frac{\nu_{\rm
        obs}}{5 {\rm GHz}}\right)^{0.35}  n_{18}^{1.24}
  \,{\rm erg \, s^{-1}} & {\rm Opt.~Thin}\\\\
  1.1 \times 10^{42} \left(\frac{E}{10^{54} {\rm ergs}}\right)^{0.87}
  \left(\frac{\epsilon_e}{0.1}\right)^{0.61}\times\\
  \left(\frac{\epsilon_b}{0.002}\right)^{0.26}\left(\frac{\nu_{\rm
        obs}}{5 {\rm GHz}}\right)^{2.01}  n_{18}^{-0.14} \,{\rm erg\,
    s^{-1}} & {\rm Opt.~Thick},
\end{dcases}
\label{eq:peakLum}
\end{align}
where we have adopted fiducial values for the power-law slope of the
gas density profile, $k=1$, and the electron energy distribution,
$p=2.3$.  The top and bottom lines apply, respectively, to the shocked
CNM being optically thin and optically thick at the deceleration time (as delineated by blue lines in Fig.~\ref{fig:diss}).

The peak luminosity in the optically thin case depends sensitively on
$n_{18}$, while in the optically thick regime the dependence on
density is much weaker.  The peak fluxes in equation
(\ref{eq:peakLum}) are normalized to match those derived from our
numerical results.

The time of maximum flux, for the same fiducial values ($k = 1$,
$p=2.3$), is given by equation (\ref{eq:tpeakGen}),
\begin{align}
&t_p =\nonumber\\ &\text{max}
\begin{dcases}
  500 E_{54}^{0.5} n_{18}^{-0.5} \,{\rm days} & {\rm Opt.~Thin}\\\\
  50 \left(\frac{E}{10^{54} {\rm ergs}}\right)^{0.32}
  \left(\frac{\epsilon_e}{0.1}\right)^{0.45}
  \left(\frac{\epsilon_b}{0.002}\right)^{0.37}\\
  \left(\frac{\nu_{\rm obs}}{5 {\rm GHz}}\right)^{-1.1} n_{18}^{0.4}
  \,{\rm days}& {\rm Opt.~Thick},
\end{dcases}
\label{eq:peakTime}
\end{align}
where again the normalizations are chosen to match our numerical
results. Note that for the optically thin case the peak time
is within a factor of two of the deceleration time
(eq.~\ref{eq:tdec}).

In general, more energetic jets produce emission which peaks later in
time.  However, the scaling of $t_p$ with $n_{18}$ is more
complicated: if the emitting region is optically thick at the
deceleration time, then the peak time increases with CNM density. In
this case the peak flux occurs when the self-absorption frequency
passes through the observing band, and this happens later if the
nuclear gas density is higher. Otherwise, peak flux is achieved near
the deceleration time, which is a decreasing function of $n_{18}$
(eq.~\ref{eq:tdec}).  Fig.~\ref{fig:diss} shows the division between
the optically-thick and optically-thin regimes at 1 and 30 GHz in the
parameter space of jet energy and $n_{18}$.

\subsection{Numerical Light Curves}
\label{sec:numResults}
As summarized in Table~\ref{tab:jetParams} (and shown in
Fig~\ref{fig:diss}), we calculate light curves for a grid of on-axis
jet simulations for five different values of $n_{18}$ (2, 11, 60, 345,
and 2000 cm$^{-3}$) and three different values of the
(beaming-corrected) jet energy $E$ ($5\times 10^{51}$, $5\times
10^{52}$, $5\times 10^{53}$ erg).

The left panels of Fig.~\ref{fig:lightcurves} show example light
curves for different jet energies and nuclear gas densities. The peak
luminosity is roughly linearly proportional to the jet energy and is
virtually independent of the ambient density.  For high CNM densities
and low frequencies this is to be expected because the emission is
dominated by the slow component, which is optically thick at the
deceleration time. However, for high frequencies and small CNM
densities, the peak luminosity of the slow component falls off, as
shown by the lighter shaded lines in the right panels of
Fig.~\ref{fig:lightcurves}. Coincidentally, the fast component just
compensates for this decline, resulting in the total (fast + slow)
on-axis peak luminosity being weakly dependent on $n_{18}$ across the entire
parameter space.  A good approximation to this universal peak
luminosity is given by equation~\ref{eq:peakLum} for $n_{18}=2000$
cm$^{-3}$ in the optically-thick case.

Fig.~\ref{fig:lightcurves} also makes clear that the peak time
increases with the ambient gas density.  Across most of the parameter
space the peak occurs after the deceleration time, when the emitting
region transitions from optically thick to optically thin, as occurs
later for larger $n_{18}$. However, at high frequencies and low
densities the slow component is optically thin at the deceleration
time, and thus its peak time is a decreasing function of $n_{18}$. For
example, at 30 GHz, the slow component peaks later for $n_{18}$=2
cm$^{-3}$ than for $n_{18}$=60 cm$^{-3}$.

\begin{figure*} 
\includegraphics[width=8cm]{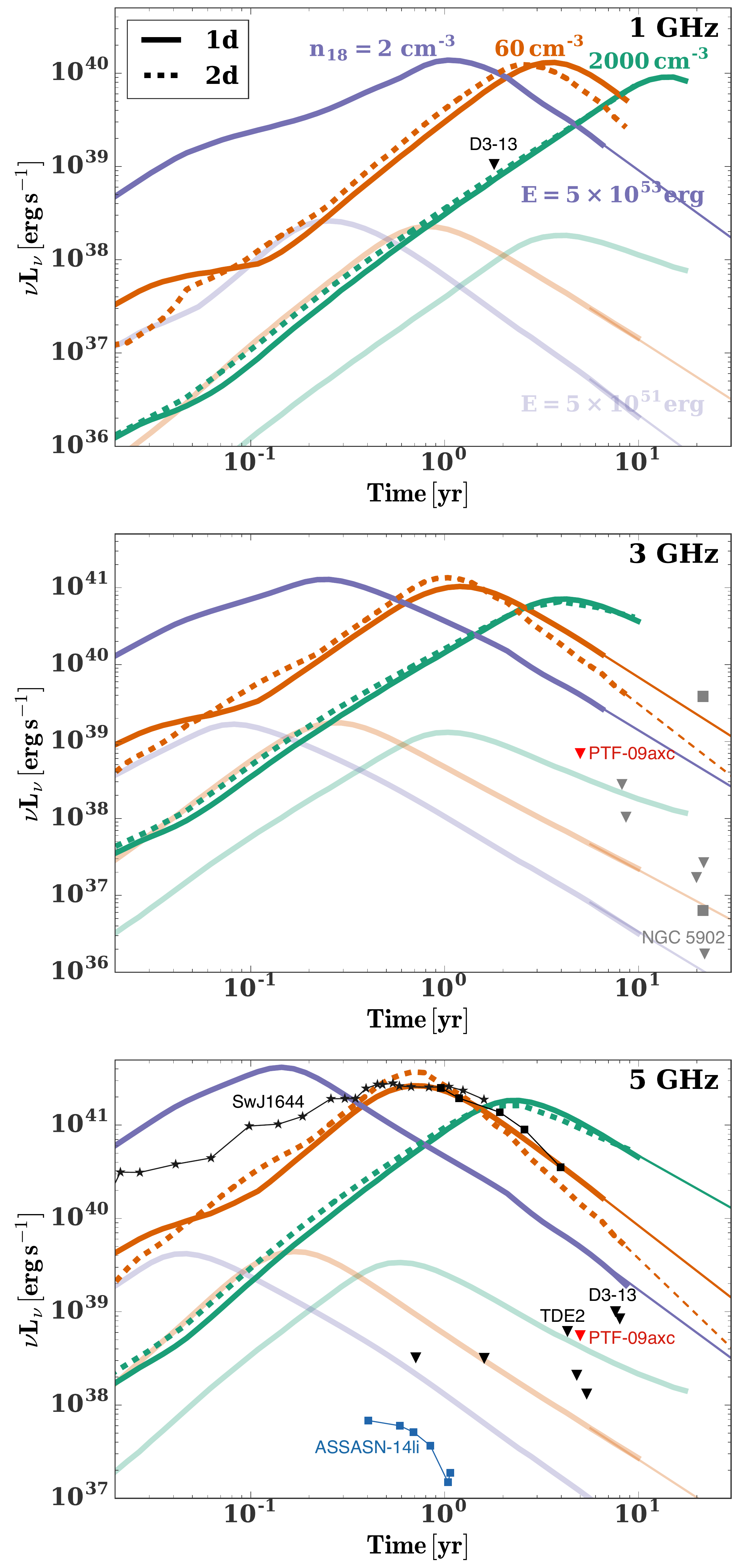}
\includegraphics[width=8cm]{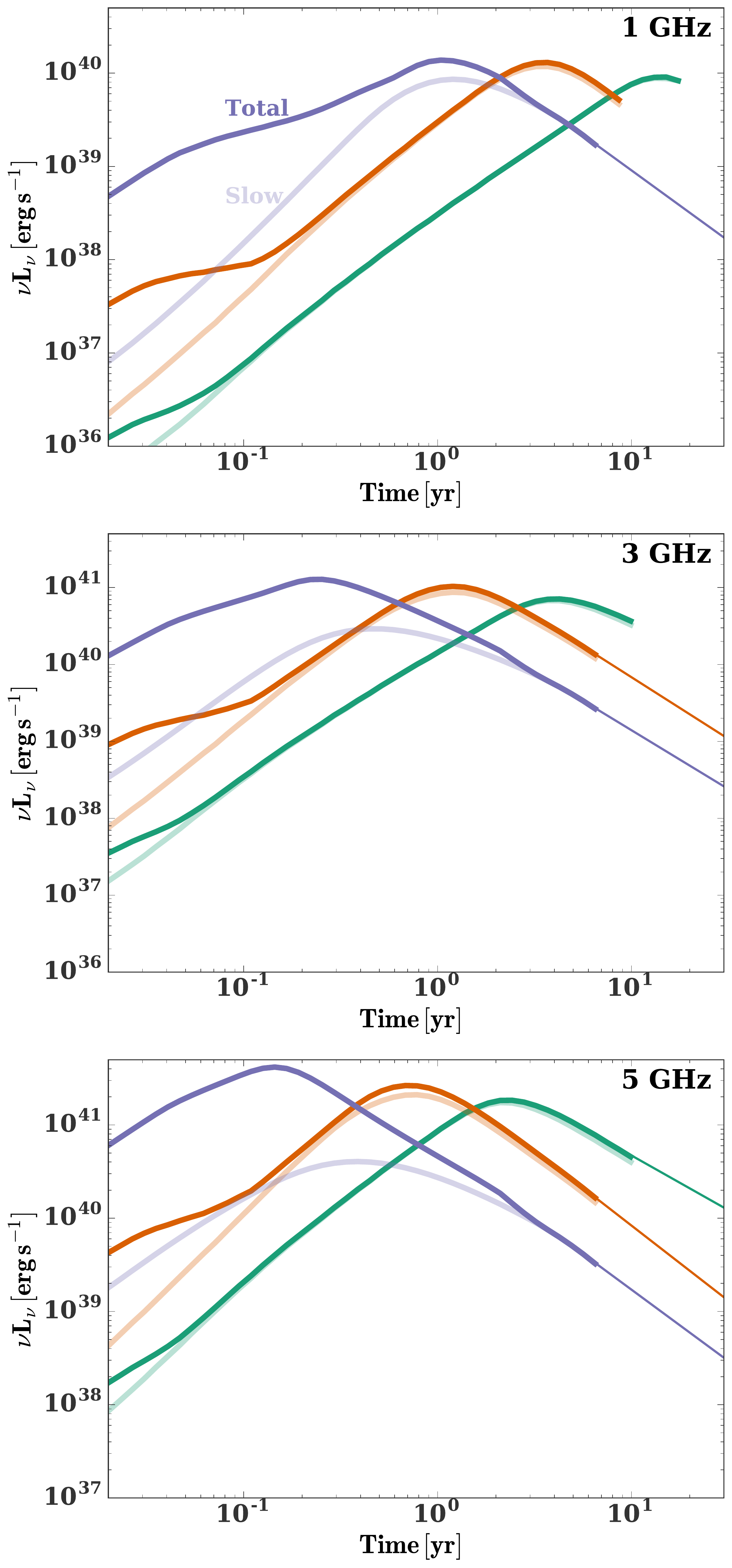}
\caption{\label{fig:lightcurves} \textit{Left:} Radio light curves as
  viewed on axis ($\theta_{\rm obs}=0$) for jet energies of $5\times
  10^{53}$ erg ({\it darker-shaded lines}) and $5\times 10^{51}$ erg
  ({\it lighter-shaded lines}), for values of n$_{18}=$ 2 (blue), 60
  (red), and 2000 (green) cm$^{-3}$.  Solid lines show the result of
  1D simulations, while 2D light curves are shown as dashed lines
  (when available). Thick lines show the results of our numerical
  calculation, while thin lines are power law extrapolations. A gas
  density profile of $n\propto r^{-1}$ is used for all of the light
  curves.  Radio upper limits and detections are shown as triangles
  and squares, respectively.  The single upper limit in the top panel
  is for D3-13 at 1.4 GHz from \citet{Bower2011}.  Gray triangles and
  squares in the second panel indicate upper limits and detections and
  detections at 3.0 GHz from \citet{Bower+2013}, while the red
  triangle is the 3.5 GHz upper limit for for PTF-09axc from
  \citet{Arcavi+2014}. Black triangles in the third panel indicate
  upper limits at 5.0 GHz from \citet{van-Velzen+2013}. The red
  triangle shows the 6.1 GHz upper limit for PTF-09axc from
  \citet{Arcavi+2014}. The connected black stars show early time data
  for SwJ1644 taken with EVLA \citep{Berger+2012, Zauderer+2013},
  while the connected black squares show late time measurement with
  the European VLBI network \citep{Yang+2016}.  Connected blue squares
  show 5 GHz data for ASSASN-14li \citep{Alexander+2016}. Note that we
  have subtracted the observed quiescent radio emission for
  ASSASN-14li). We have labeled events which have upper limits across
  multiple frequencies \textit{Right}: $5\times 10^{53}$ erg on-axis
  light curves from left column ({\it darker-shaded lines}) and
  corresponding slow component light curves ({\it lighter-shaded
    lines}). Figure is continued on next page.}
\end{figure*}

\begin{figure*}
  \contcaption{Simulation results at 8 and 30 GHz. Top left panel
    includes 8.4 GHz and 7.9 GHz upper limits for TDE2 and
    SDSSJ1201+30 respectively (see Table~\ref{tab:enConstr})}
  \includegraphics[width=8cm]{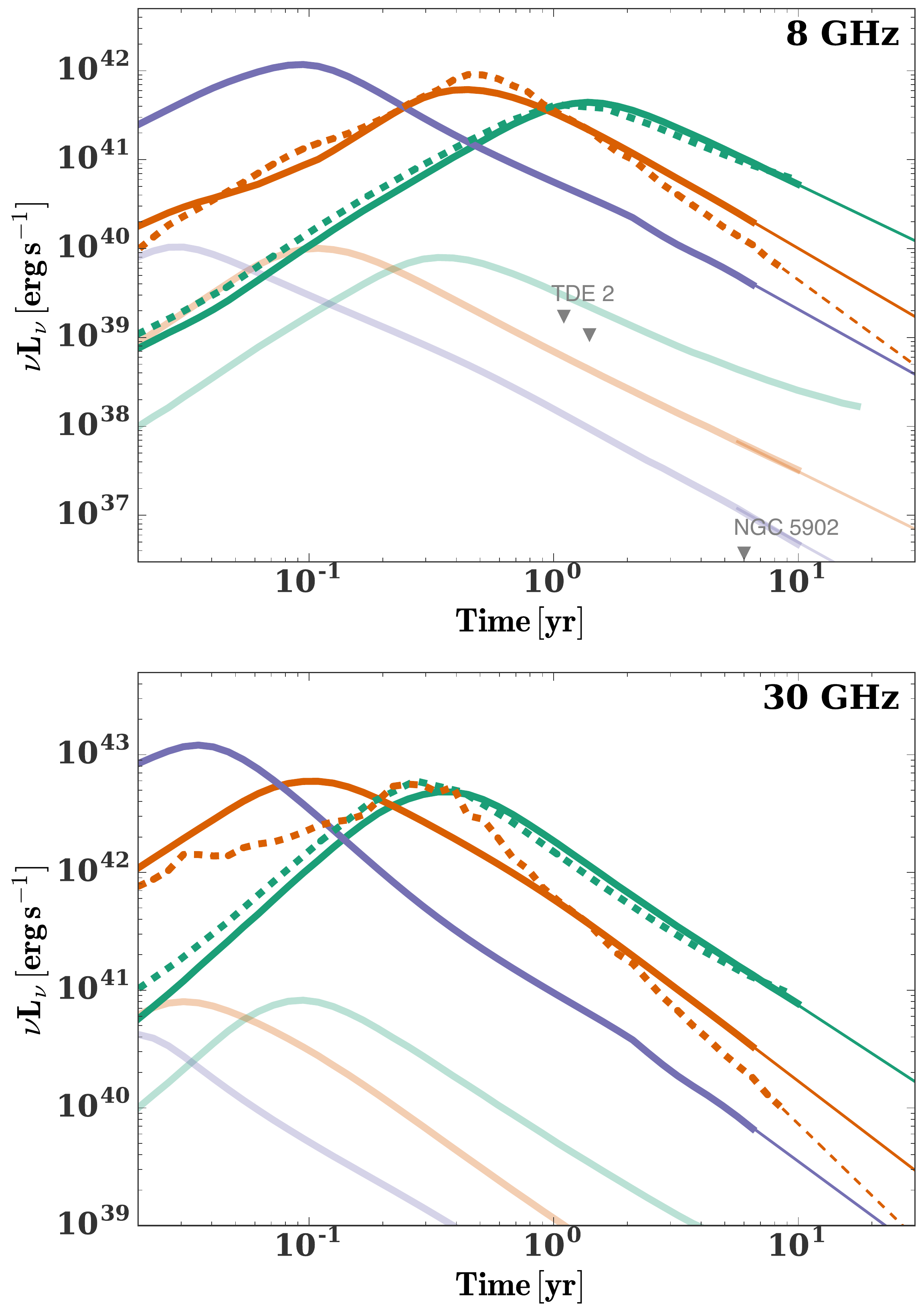}
  \includegraphics[width=8cm]{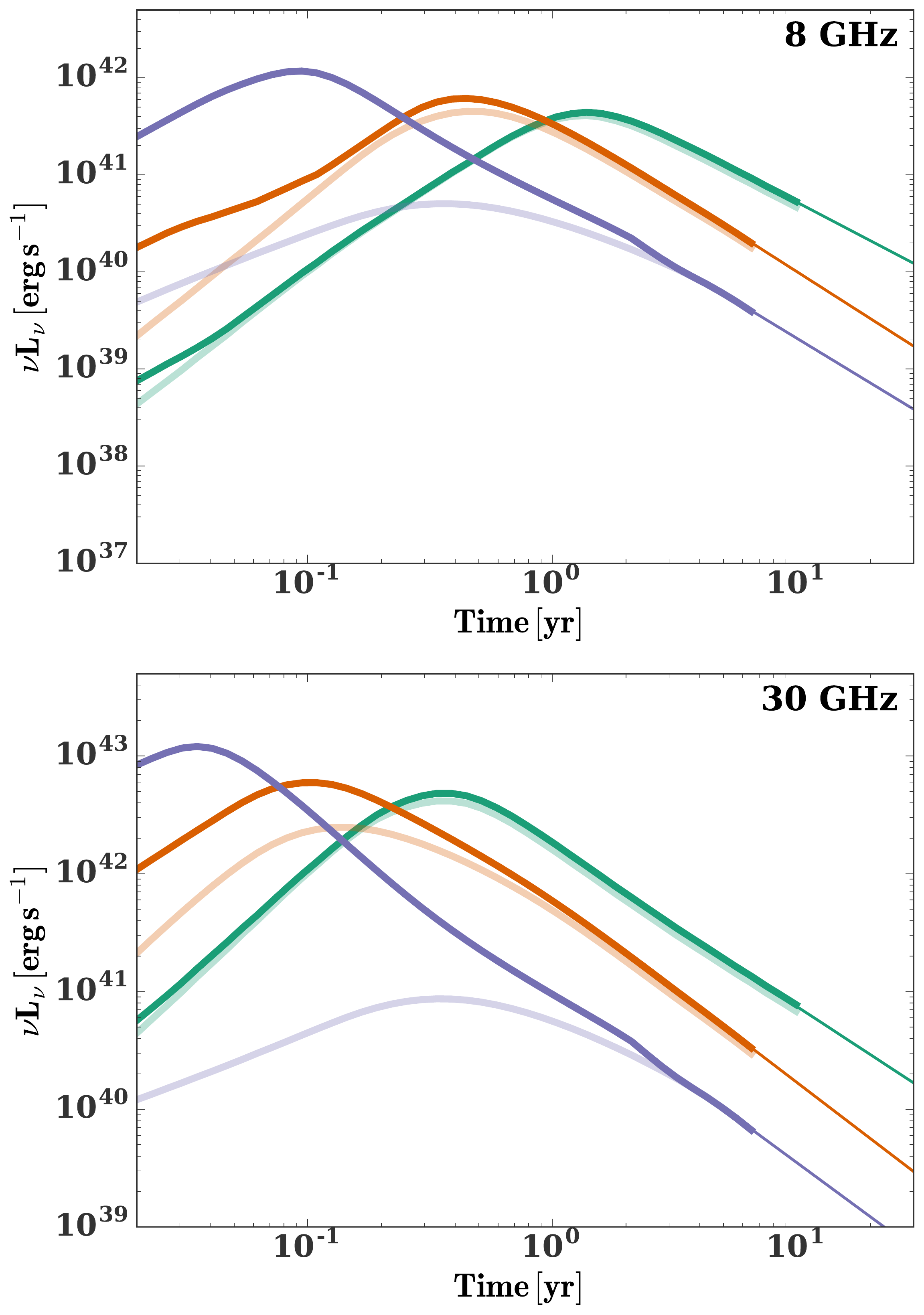}
\end{figure*}

The numerical light curves are well fit by a broken power law (see
e.g.~\citealt{Leventis+2012}),
\begin{equation}
L_\nu (t) =\frac{L_{\nu, p}}{2^{-1/s}}
\left[\left(\frac{t}{t_p}\right)^{-s
    a_1}+\left(\frac{t}{t_p}\right)^{-s a_2}\right]^{-1/s}, 
\label{eq:lcAnal}\end{equation}
where $L_{\nu, p}$ and $t_p$ are the peak luminosity and time given by
equations~\eqref{eq:peakLum} and~\eqref{eq:peakTime}, respectively.
The parameter $s$ controls the sharpness of the transition between the
early-time power-law slope $a_1$ and the late-time slope $a_2$.
Fitting to the numerical light curves, we find that $s\sim 1.0$,
$a_1\sim 1.7$, and $a_2\sim -1.4$, the latter approximately agreeing
with the analytic estimate in equation~\eqref{eq:tslope}. These
parameters generally reproduce our numerical light curves to within a
factor of a few throughout our parameter space. However, the highest
density/lowest energy light curve diverges from the power law fit at
late times as the outflow enters into the deep Newtonian regime (see
\citealt{Sironi&Giannios2013}). Also, the 2D, $n_{18}=60$ cm$^{-3}$
light curve has a somewhat steeper late time light curve
that declines as $t^{-2}$.

\begin{figure}
\includegraphics[width=8cm]{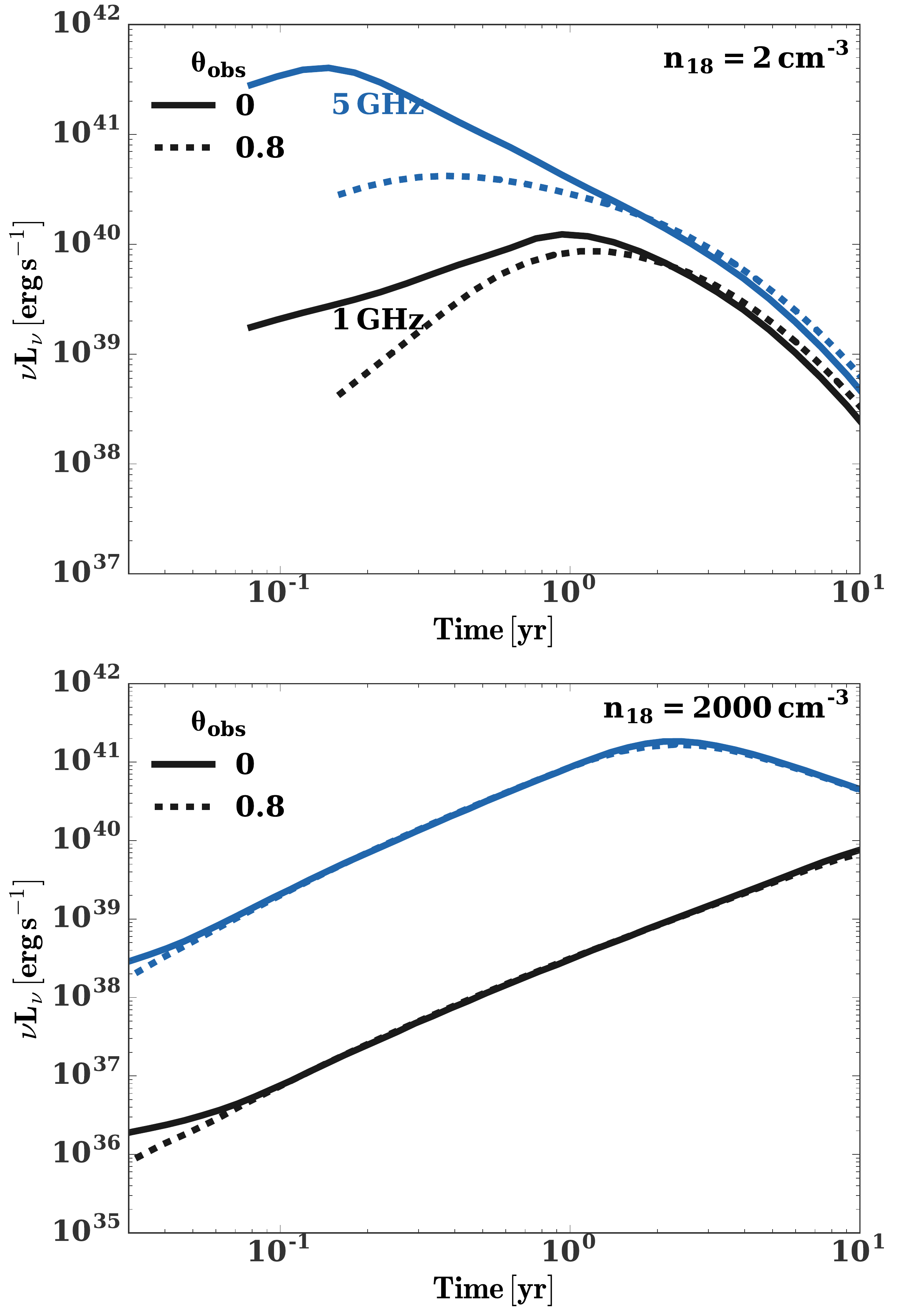}
\caption{\label{fig:onOff} Comparison between on-axis (solid line) and
  off-axis (dashed line) light curves from our 1D simulations.  The
  off-axis light curves are calculated for an observer viewing angle
  of $\theta_{\rm obs}$=0.8.  We adopt a density profile of $n\propto
  r^{-1}$. We note that the steepening of the $n_{18}=2$ cm$^{-3}$ light
  curves after 2 years is not physical and is due to limited angular
  resolution (see \citealt{Mimica+2016}).}
\end{figure}

Fig.~\ref{fig:onOff} compares the light curves for observers aligned
with the jet axis (on-axis) with those at an angle of 0.8 radians from
the jet axis (off-axis).  While the on- and off-axis light curves
agree well for $n_{18}=2000$ cm$^{-3}$, the off-axis luminosity for
$n_{18}=2$ cm$^{-3}$ is smaller by an order of magnitude at peak.
This is because the peak of the on-axis light curve is
dominated by the fast component of the jet, which would not be visible
for significantly off-axis observers. However, we find that the late
time light curve is nearly independent of viewing angle.

\begin{figure} 
  \includegraphics[width=8cm]{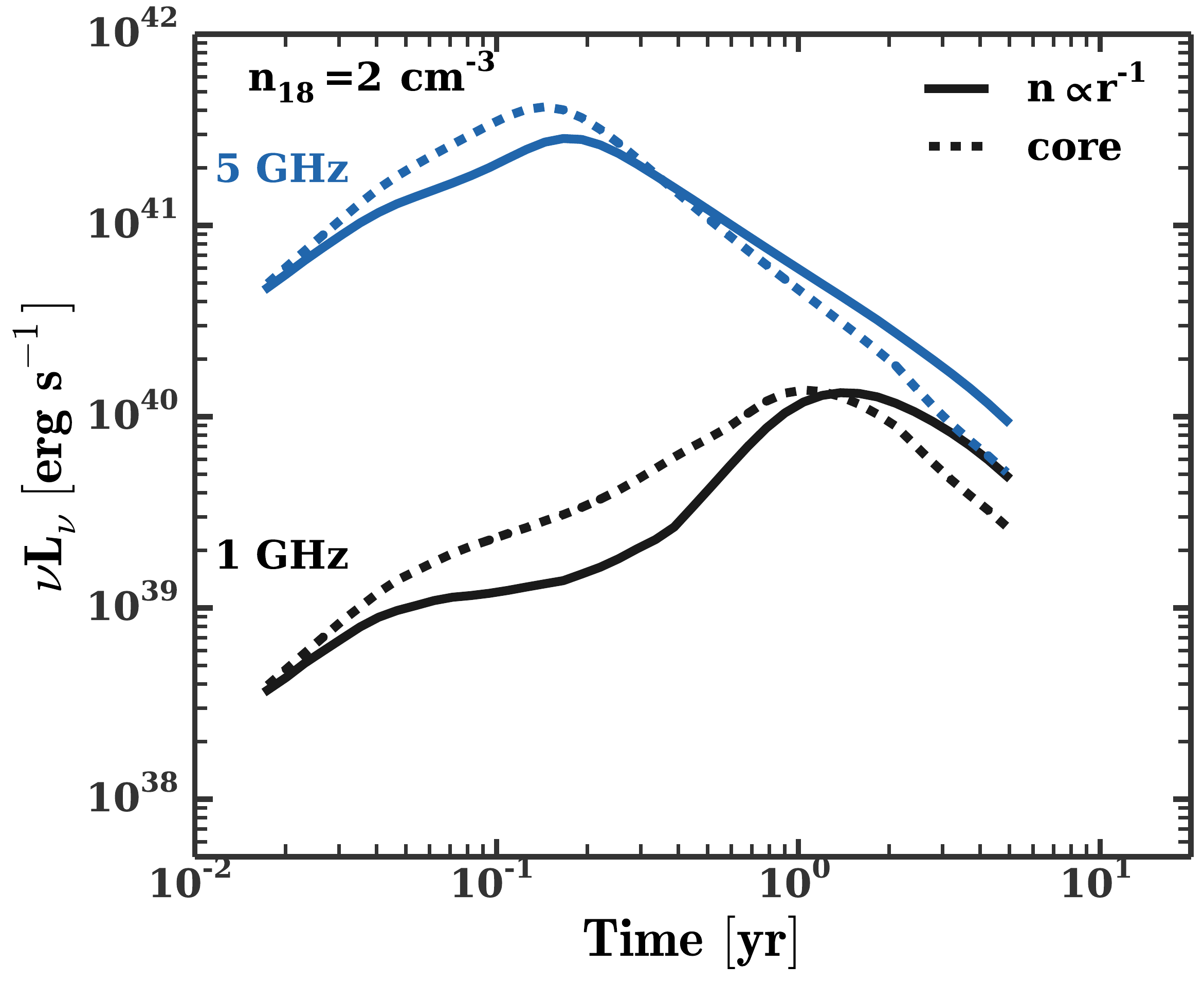}
  \includegraphics[width=8cm]{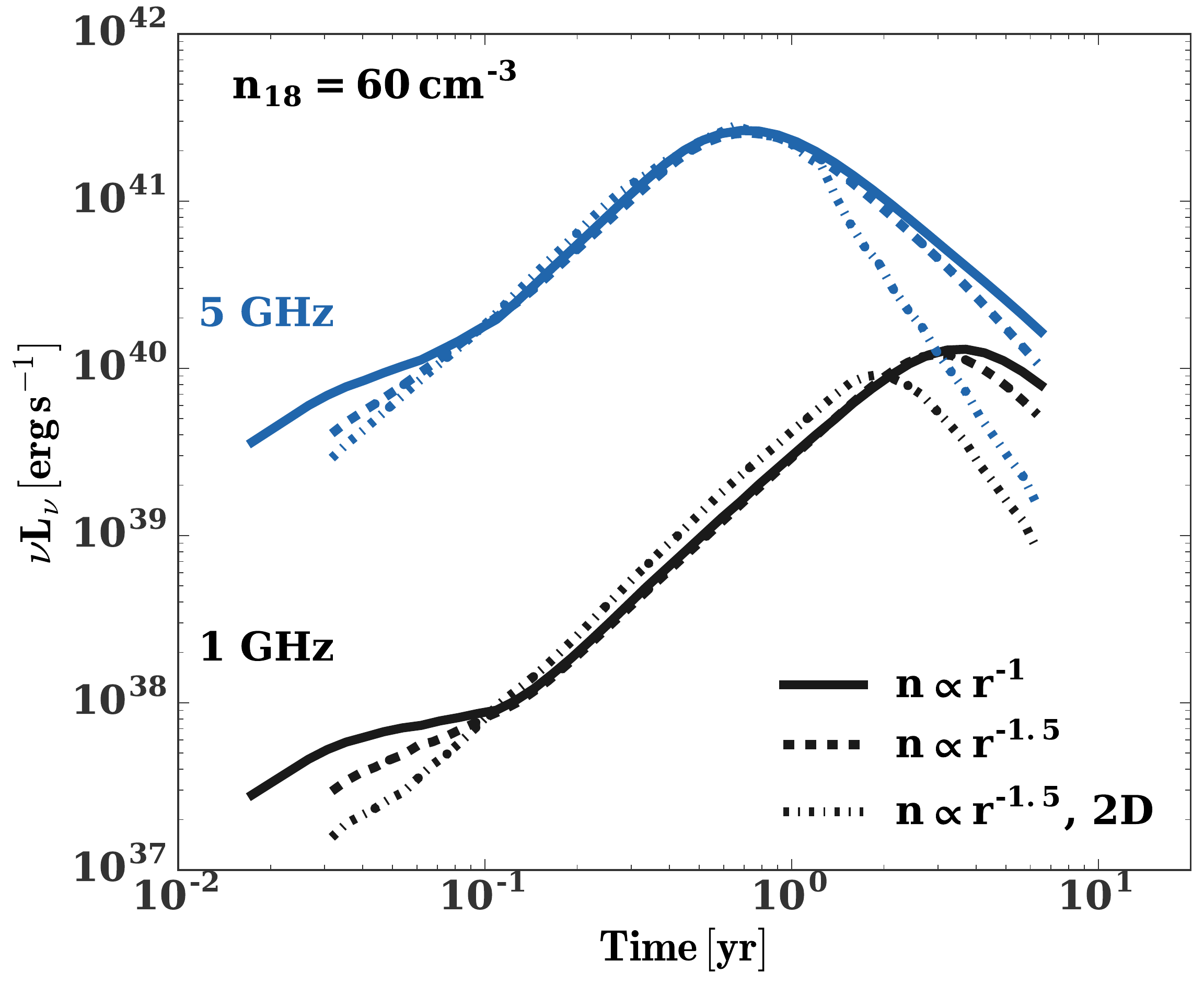}
  \caption{\label{fig:cores} {\it Top:} Comparison between on-axis
    light curves for our fiducial $n\propto r^{-1}$ gas density
    profile, corresponding to a cusp-like galaxy, and the core galaxy
    profile defined by \eqref{eq:cores} with $r_s=10^{18}$ cm. {\it
      Bottom:} Comparison between on-axis light curves calculated from
    1D simulations with $n\propto r^{-1}$ ({\it solid}) and $n\propto
    r^{-1.5}$ ({\it dashed}) gas density profiles. The dash-dotted line
    shows the on-axis light curve for a 2D simulation with an
    $n\propto r^{-1.5}$ gas density profile.}
\end{figure}

\label{sec:profileComp}
The top panel of Fig.~\ref{fig:cores} shows 1D on-axis radio light
curves for our fiducial gas density profile, $n\propto r^{-1}$, and a
core galaxy profile (equation~\ref{eq:cores}), both with $n_{18}=2$
cm$^{-3}$.  The light curves differ by at most a factor of a few. The
core and cusp light curves are even closer at higher densities, and
virtually indistinguishable at $n_{18}=2000$ cm$^{-3}$. This is
because for larger ambient densities, the jet only samples small
radii, where the core and cusp profiles are similar (see
Fig.~\ref{fig:profiles}). It is only at lower densities, for which the
Sedov radius lies outside of the flattening of the core density
profile, that noticeable differences emerge.

The bottom panel of Fig.~\ref{fig:cores} compares the 1D on-axis light
curves for $n\propto r^{-1}$ and $n\propto r^{-1.5}$ gas density
profiles with $n_{18}=60$ cm$^{-3}$. For most times the light curves
agree well, which is perhaps not surprising because the density in
these two models agrees at $10^{18}$ cm, which is close to the Sedov
radius for these density profiles. However, In 2D hydrodynamical
simulations, a jet propagating through an $r^{-1.5}$ density profile
develops a more prolate structure than a jet propagating through an
$r^{-1}$ profile. This results in a light curve with a much steeper
late time slope (see dash-dotted line in Fig.~\ref{fig:cores}),
although we note that the peak luminosity is nearly the same for the
$n\propto r^{-1}$ and $n\propto r^{-1.5}$ density profiles.

\subsubsection{Reverse Shock Emission?}
Our calculations shown in Figs.~\ref{fig:1D2DB}
and~\ref{fig:lightcurves}-\ref{fig:cores} include only emission from
the forward shock (shocked CNM), while in principle the reverse shock
(shocked jet) also contributes to the radio light curve.

The fraction of the initial kinetic energy of the jet which is
dissipated by the reverse shock provides a first-order estimate of its
maximum contribution to the radio light curve.  Fig.~\ref{fig:diss}
shows contours of the fraction of the kinetic energy of the slow
component dissipated by the reverse shock as a function of the jet
energy and CNM density, $n_{18}$.  This is estimated by integrating
the shock evolution determined from the jump conditions (see
Appendix~\ref{sec:reverse} for details), approximating the jet as a
constant source of duration $t_0 = 5 \times 10^{5}$ s and Lorentz
factor $\Gamma = 2$.  The parameters defining our grid of numerical
solutions are shown in Fig.~\ref{fig:diss} as red squares.

\begin{figure}
  \includegraphics[width=8cm]{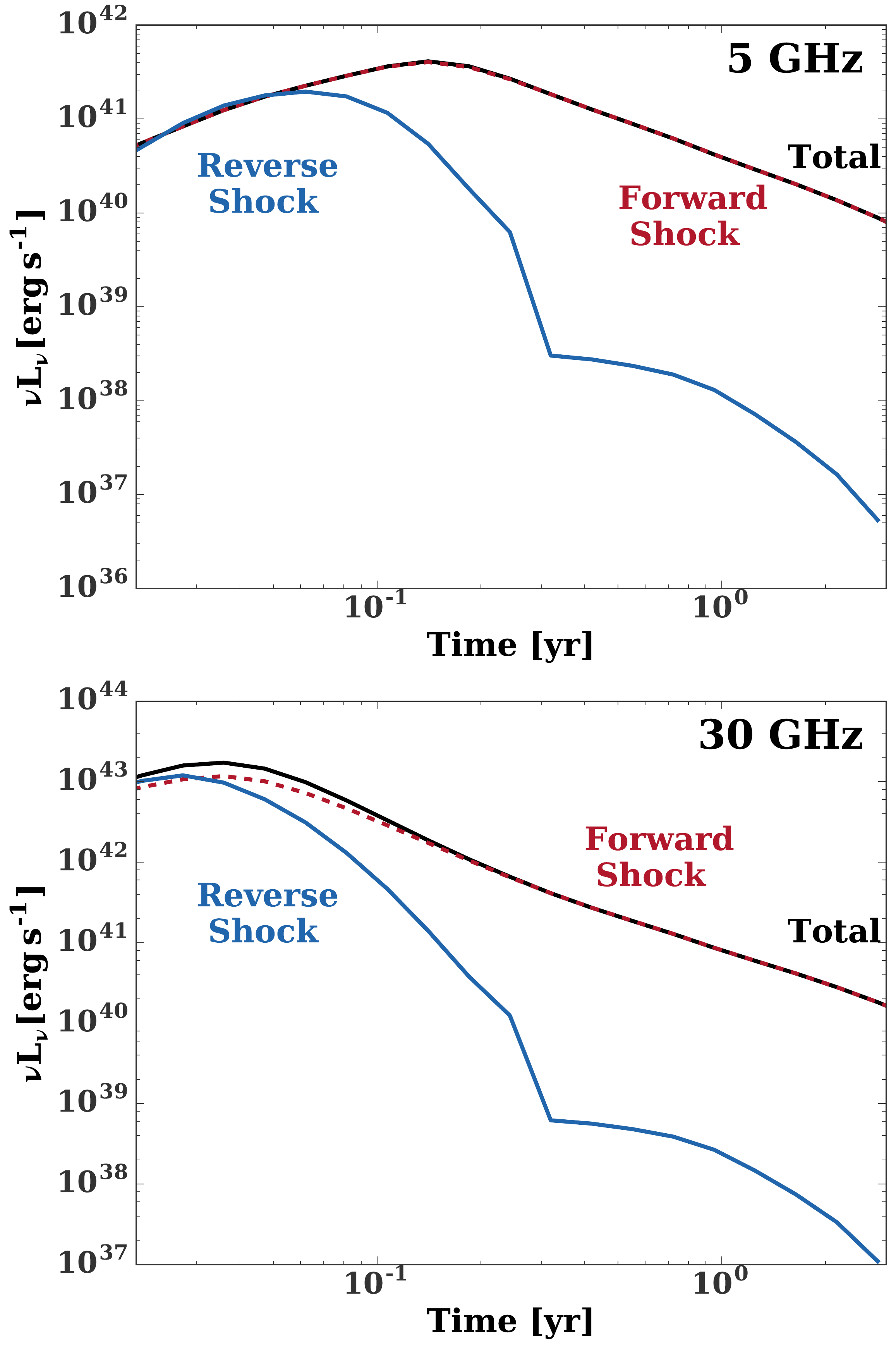}
  \caption{\label{fig:reverse} Radio light curve from the forward
    shock (red line), reverse shock (blue), and the total light curve
    (black) for a jet of energy $5\times 10^{53}$ erg and CNM density
    $n\propto r^{-1}$ with $n_{18} = 2$ cm$^{-3}$. The reverse shock
    light curve excludes absorption from the front of the jet, which
    when included in the full calculation results in large attenuation
    of the emission, such that the total light curve is dominated by
    the forward shock.}
\end{figure}

Fig.~\ref{fig:diss} shows that for high ambient densities and/or low
energy jets, the reverse shock dissipates an order unity fraction of
the kinetic energy of the jet.  Even for our highest energy/lowest
density model ($n_{18}=2$ cm$^{-3}$ and $E_j=5\times 10^{53}$ erg) the
reverse shock will dissipate of order $~20\%$ of the jet energy.
Fig.~\ref{fig:reverse} shows the 5 GHz and 30 GHz light curve for this
case, separated into contributions from the forward and reverse
shocks.  The reverse shock emission is comparable to that from the
forward shock for the first month.  However, this overstates the true
contribution of the reverse shock to the observed emission because the
latter is strongly attenuated by absorption from the front of the jet,
which has not been included in the reverse shock light curve in
Fig.~\ref{fig:reverse}. For 5 GHz the contribution of the reverse
shock to the total light curve is negligible at all times. For the 30
GHz, the peak luminosity increases by a factor of $1.5$ after reverse
shock emission is taken into account.  While the reverse shock
dissipates an even larger fraction of the jet energy for higher
ambient density, its emission will be even more heavily absorbed.  We
conclude that the reverse shock emission can be neglected for the high
energy jets with $E\gtrsim 10^{53}$ erg, consistent with the reverse
shock not contributing appreciably to SwJ1644
(\citealt{Metzger+2012}).

For low energy jets, we find that the jet is crushed at early times,
even for low values of $n_{18}$. In the case of very low power jets
the reverse shock structure is replaced by a number of recollimation
shocks (similar to the structure seen in
e.g. \citealt{Mimica+2009}). While this is potentially a very
interesting case since the emitting volume from recollimation shocks
can be larger than from a single reverse shock, because of a much more
complex structure we defer a more detailed study of the emission from
the reverse/recollimation shocks in the the low energy case to future
work.

As a final note of caution, even if the reverse shock dissipates most
of the bulk kinetic energy into thermal energy, the latter can be
converted back to kinetic energy through adiabatic expansion.
However, we expect that the re-expansion will be relatively isotropic
compared to the original jet, because the matter is first slowed to
mildly relativistic speeds.  The net result of a ultra-strong reverse
shock (due to a weak jet, and/or an unusually high CNM density) is
therefore likely to be the production of two quasi-spherical lobes on
either side of the black hole, centered about the deceleration radius
(\citealt{Giannios&Metzger2011}).

\subsection{Parameter Space of Jet-CNM Interaction}
\label{sec:param}
\begin{figure*}
  \includegraphics[width=16cm]{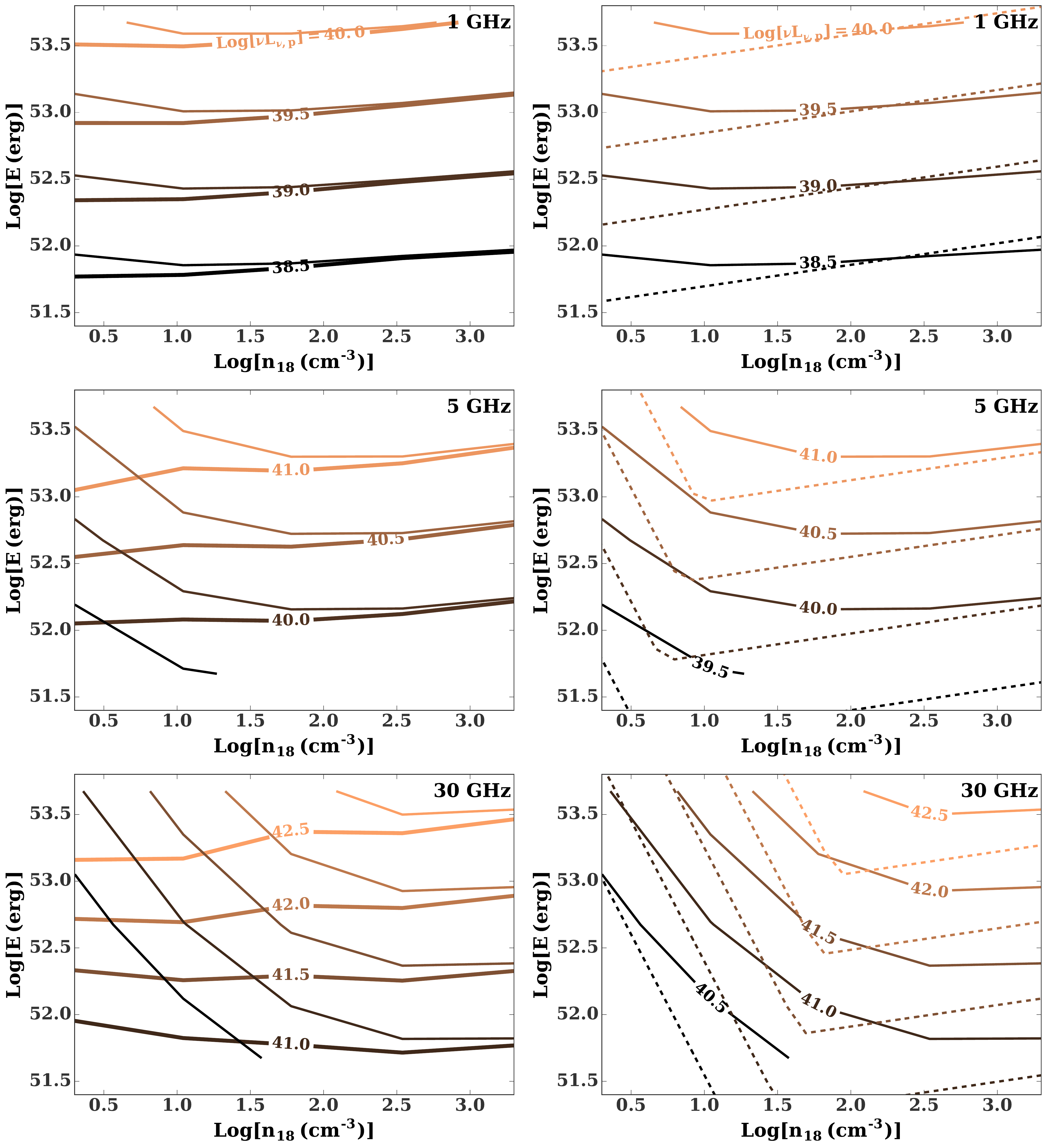}
  \caption{\label{fig:jetContours} {\it {Left:}} Thick lines show the
    peak radio luminosity in the parameter space of jet energy and
    ambient gas density at $10^{18}$ cm, calculated from the grid of
    on-axis jet simulations in Table~\ref{tab:jetParams}. Thin lines
    show contours of peak luminosity for the slow component light
    curve ($\S$~\ref{sec:numerical}). {\it Right:} Analytic estimate
    for the peak luminosity (dashed lines; eq.~\ref{eq:peakLum})
    compared to the numerical results for the slow component (solid
    lines).}
\end{figure*}

The left column of Fig.~\ref{fig:jetContours} shows contours of the
peak luminosity (thick lines) as derived from our grid of numerical
on-axis models, covering the parameter space of jet energy $E$ and
density $n_{18}$.  Also shown with thin lines is the luminosity
arising from just the slow, wide angle component.  The fast, narrow
component of the jet dominates at high frequencies and low densities,
while the slow, wide component dominates for large $n_{18}$ and low
frequencies.  Remarkably, the total peak luminosity is nearly
independent of the ambient gas density; this is in part coincidental,
as the fast and slow peak fluxes individually vary across the
parameter space. For off-axis jets, the peak luminosity is dominated
by just that of the slow component, and thus would be a decreasing
function of the ambient density above 1 GHz.

The right column of Fig.~\ref{fig:jetContours} compares our numerical
results for the slow component to the analytic estimate given in
equation \eqref{eq:peakLum}.  For large $n_{18}$, the optically thick
case reproduces the peak luminosity to within a factor of a few.  By
contrast, for 30 GHz and low $n_{18}$, the numerical results are
closer to the optically thin limit.

\begin{figure*}
  \includegraphics[width=16cm]{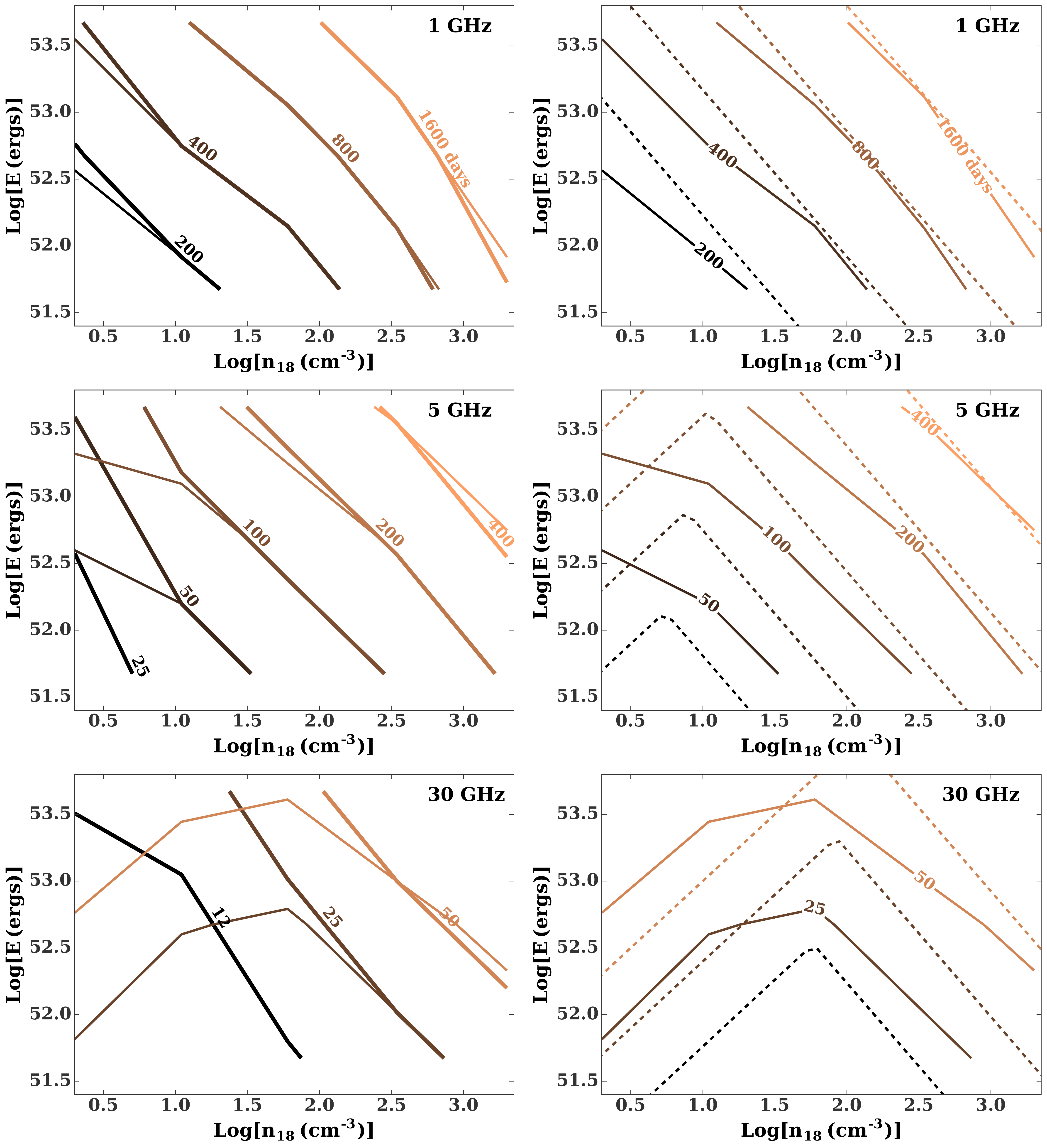}
  \caption{\label{fig:ContoursTp} {\it {Left:}} Thick lines show peak
    time in days in the parameter space of jet energy and ambient gas
    density at $10^{18}$ cm, calculated from the grid of on-axis jet
    simulations in Table~\ref{tab:jetParams}. Thin lines show contours
    of peak time for the slow component light curve
    (see~\ref{sec:numerical}). {\it Right:} Analytic scaling for the
    peak time ({\it dashed}, see equation~\ref{eq:peakTime}) compared
    to the numerical results for the slow component (solid)}
\end{figure*}

The left column of Fig.~\ref{fig:ContoursTp} shows contours of the
time of peak flux in days, separately for the slow component (thin
lines) and the total light curve (thick lines).  Shown for comparison
in the panels in the right column is the peak time as estimated from
equation~\eqref{eq:peakTime}. At 30 GHz, the peak time decreases with
$n_{18}$ at small values of the latter, because in this regime the jet
is optically thin prior to the deceleration time.

\subsubsection{Comparison with radio detections and upper limits.}
\label{sec:upLims}

Fig.~\ref{fig:lightcurves} compares our fiducial $5\times 10^{53}$ erg
on-axis jet model to radio detections and upper limits derived from
follow-up observations of TDE flares (including
SwJ1644\footnote{Detailed comparison of our model with radio data from
  SwJ1644 data is given in \citealt{Mimica+2015}.}), as compiled in
Table~\ref{tab:enConstr}.  All of the 5 GHz light curves,
corresponding CNM densities, $n_{18}$, of 2, 60, and 2000 cm$^{-3}$,
fall above the upper limits. In agreement with the results of previous
work, we conclude that most TDEs discovered by their optical/UV or
soft X-ray emission do not produce jets as powerful as that
responsible for SwJ1644
(\citealt{Bower+2013,van-Velzen+2013,Mimica+2015}), a result which is
now found to hold for a broad range of CNM environments.

The peak radio luminosity at frequencies $\lsim$ 1 GHz is weakly
dependent on the ambient gas density. Radio observations conducted
from several months to years after a tidal disruption flare, which
tightly constrain the peak flux of a putative jet, can therefore be
used to constrain the jet energy.  Equation~\eqref{eq:peakLum} shows
that an upper limit of $F_{\rm ul}$ on the flux density at 1 GHz of a
source at distance $d_L$ results in an upper limit on the jet energy
of
\begin{equation}
  E\lsim 4.3 \times 10^{49} \left(\frac{F_{\rm ul}}{50 \,\mu{\rm Jy}}\right)^{1.1}
  \left(\frac{d_L}{200 \,{\rm Mpc}}\right)^{2.3} {\rm erg},
\label{eq:upLim}
\end{equation}
where we have taken $n_{18}=2000$ cm$^{-3}$ (but the constraint is not
overly sensitive to this choice for $n_{18}\geq 2$
cm$^{-3}$)\footnote{The peak luminosity will decrease approximately
  linearly with $n_{18}$ for $n_{18}\leq 2$ cm$^{-3}$.  For the
  smallest plausible value of $n_{18}$, 0.3 cm$^{-3}$, the
  normalization in equation~\eqref{eq:upLim} would increase by a
  factor of 7.}.  Radio measurements of the peak flux following a TDE
therefore serve as calorimeters of the total energy released in a
relativistic jet (or spherical outflow).

\begin{figure}
\includegraphics[width=8.5cm]{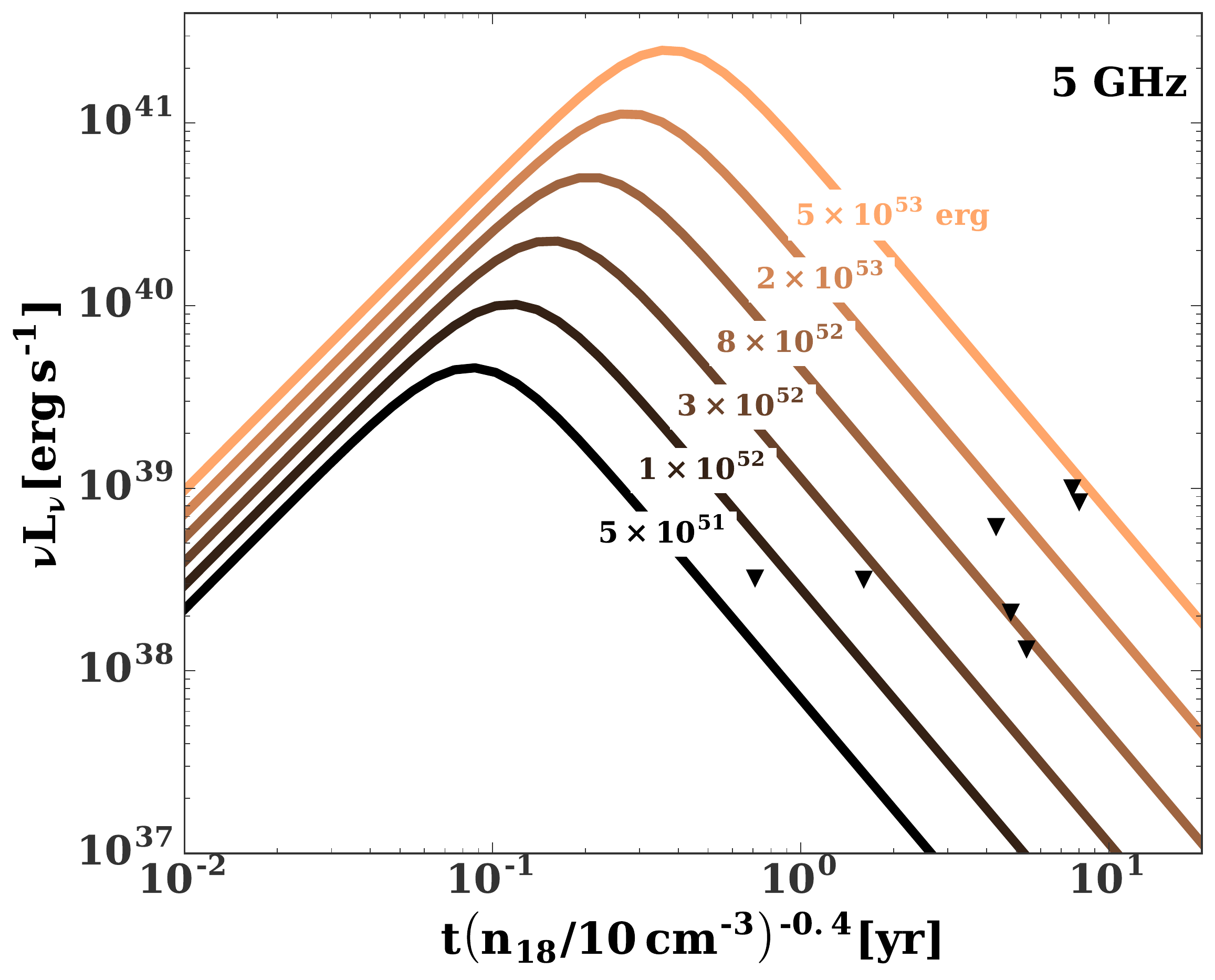}
\caption{\label{fig:econtours} 
Upper limits and 5 GHz analytic light
  curves (eqn.~\ref{eq:lcAnal} with $s=1$, $a_1=1.7$, and $a_2=-2$)
  for different jet energies.  We use the peak time and luminosity
  from our numerical $n_{18}$=11 cm$^{-3}$ light curve for the highest
  energy light curve, as our analytic fits (eqns.~\ref{eq:peakLum}
  and~\ref{eq:peakTime}) underestimate the peak luminosity a factor of
  $\sim$2 for this density. Then we use our analytic results to scale
  this light curve to lower energies. 
 }
\end{figure}

If the peak flux is missed, late time measurements can still be used
to constrain the jet energy. In fact, with late time measurements it
is possible to place constraints on the energy of the jet/outflow
using higher frequency radio data. Fig.~\ref{fig:econtours} compares
our analytic fit to the on-axis 5 GHz synchrotron light curve
(eq.~\ref{eq:lcAnal}) for different jet energies and existing radio
upper limits for $n_{18}$=10 cm$^{-3}$, the minimum expected density
for stellar populations observed in TDE host galaxies.  An increase in
$n_{18}$ would simply shift the light curves to the right. Thus, for
times after peak each light curve in Fig.~\ref{fig:econtours} gives
smallest plausible radio luminosity for the corresponding jet
energy. As the upper limits are all taken at late times, the
$n_{18}$=10 cm$^{-3}$ light curve which passes through each upper
limit corresponds to the maximum jet energy consistent with it. We
note that that in this case, the deceleration radius is inside both
the influence radius and the stagnation radius, and thus we would
expect the density profile there to be closer to $r^{-1.5}$, rather
than $r^{-1}$. A steeper density profile would cause a steeper late
time decline in the light curve, and would make the upper limits less
constraining.  However, the steeper profile would imply a larger
density at $n_{18}$, which would compensate for this.

\begin{figure}
\includegraphics[width=8.5cm]{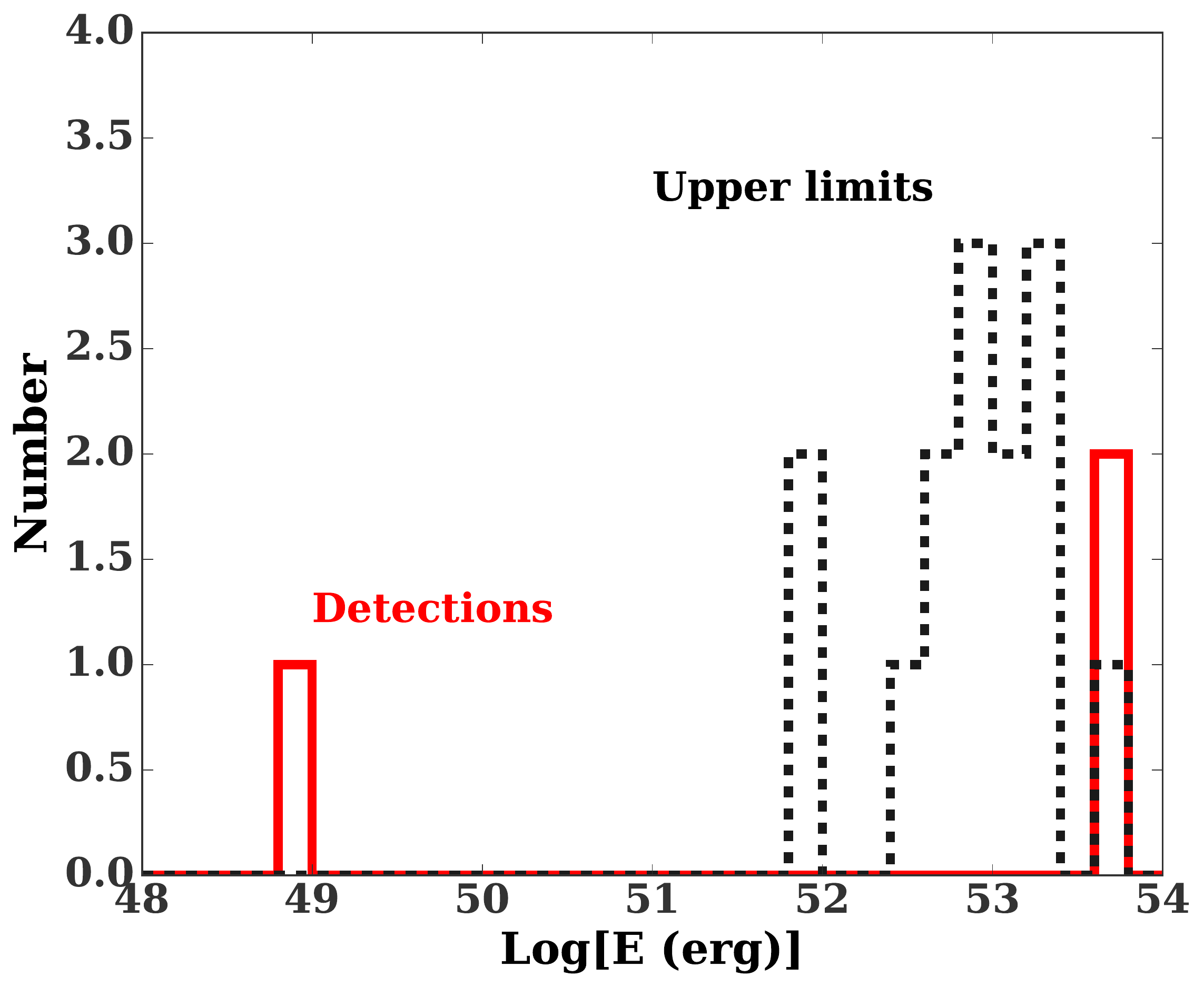}
\caption{\label{fig:hist} Histogram of jet energies consistent with
  existing radio detections (ASSASN-14li,  SwJ1644, and
  SwJ2058) and upper limits (Table 1 of \citealt{Mimica+2015} and
  \citealt{Arcavi+2014}), as summarized in Table~\ref{tab:enConstr}.}
\end{figure}

Fig.~\ref{fig:hist} shows a histogram of the maximum jet energies
consistent with the existing radio upper limits and detections of TDE
flares with radio follow-up (see also Table \ref{tab:enConstr}).  The
detected events include ASSASN-14li, SwJ1644, and SwJ2058.  For
ASSASN-14li and SwJ1644 the lightcurves are well sampled, and the
energy of the jet is relatively well constrained to be $\approx
10^{48}-10^{49}$ erg for ASSASN-14li (\citealt{van-Velzen+2016,
  Alexander+2016}) and $5\times 10^{53}$ erg for SwJ1644
(\citealt{Mimica+2015}).  For SwJ2058, we take the jet energy to be
$5\times 10^{53}$ erg, the same as its ``twin'' SwJ1644 \citep{Cenko+2012,Pasham+2015}.

\begin{table*}
\begin{threeparttable}
  \caption{\label{tab:enConstr} Inferred jet/outflow energies (and
    bounds) from radio detections and upper limits of optical/UV and
    soft X-ray TDE candidates. For each event detected in the radio
    there are multiple observations at different
    times/frequencies. Thus, we leave a dash in the time frequency,
    and luminosity columns and simply to refer to reference in column
    ``Ref.''}
\begin{tabular*}{1.5\columnwidth}{lllllll}
  \hline
  Source & $D_L$ & t & $\nu$ & $\nu L_{\nu}$ & Ref. & Energy\\
  & (Mpc) & (yr) & (GHz) & ($10^{36}$ erg s$^{-1}$) & & (erg) \\
  \hline
  Detections \\
  \hline
  ASSASN-14li & 93 & - & - & - & 1 &  $10^{48}-10^{49}$\\
  SwJ1644 & 1900 & - & - & - &  2  & $5\times 10^{53}$\\
  SwJ2058 & 8400  & - & - & - & 3 & $5\times 10^{53}$\\ 
  \hline 
  Upper limits & \\
  \hline
  RXJ1624+7554 & 290 & 21.67 & 3.0 & 27 & 4 & $< 1.4 \times 10^{ 53 }$ \\
RXJ1242-1119 & 230 & 19.89 & 3.0 & 17 & 4 & $< 9.6 \times 10^{ 52 }$ \\
SDSSJ1323+48 & 410 & 8.61 & 3.0 & 100 & 4 & $< 1.0 \times 10^{ 53 }$ \\
SDSSJ1311-01 & 900 & 8.21 & 3.0 & 280 & 4 & $< 1.9 \times 10^{ 53 }$ \\
D1-9 & 1800 & 8.0 & 5.0 & 840 & 5 & $< 4.1 \times 10^{ 53 }$ \\
TDE1 & 660 & 5.4 & 5.0 & 130 & 5 & $< 7.1 \times 10^{ 52 }$ \\
D23H-1 & 930 & 4.8 & 5.0 & 210 & 5 & $< 8.2 \times 10^{ 52 }$ \\
PTF10iya & 1100 & 1.6 & 5.0 & 320 & 5 & $< 2.5 \times 10^{ 52 }$ \\
PS1-10jh & 840 & 0.71 & 5.0 & 320 & 5 & $< 8.7 \times 10^{ 51 }$ \\
NGC5905 & 49 & 21.91 & 3.0 & 1.7 & 4 & $< 2.4 \times 10^{ 52 }$ \\
NGC5905 & 49 & 6.0 & 8.6 & 3.7 & 6 & $< 8.2 \times 10^{ 51 }$ \\
D3-13 & 2000 & 7.6 & 5.0 & 1000 & 5 & $< 4.3 \times 10^{ 53 }$ \\
D3-13 & 2000 & 1.8 & 1.4 & 1000 & 7 & $< 2.5 \times 10^{ 53 }$ \\
TDE2 & 1300 & 4.3 & 5.0 & 610 & 5 & $< 1.4 \times 10^{ 53 }$ \\
TDE2 & 1300 & 1.1 & 8.4 & 1700 & 8 & $< 5.0 \times 10^{ 52 }$ \\
SDSSJ1201+30 & 710 & 1.4 & 7.9 & 1100 & 9 & $< 5.0 \times 10^{ 52 }$ \\
PTF09axc & 550 & 5.0 & 3.5 & 700 & 10 & $< 1.8 \times 10^{ 53 }$ \\
PTF09axc & 550 & 5.0 & 6.1 & 550 & 10 & $< 1.7 \times 10^{ 53 }$ \\
\end{tabular*}
\begin{tablenotes}
\item References: $(1)$ \citet{Alexander+2016, van-Velzen+2016}, $(2)$
  \citet{Berger+2012, Zauderer+2013, Yang+2016} $(3)$
  \citet{Cenko+2012}, $(4)$ \citet{Bower+2013}, $(5)$
  \citet{van-Velzen+2013}, $(6)$ \citet{Bade+1996,
    Komossa&Dahlem2001}, $(7)$ \citet{Gezari+2008,Bower+2011}, $(8)$
  \citet{van-Velzen+2011}, $(9)$ \citet{Saxton+2012}, $(10)$
  \citet{Arcavi+2014}. All upper limits are 5 $\sigma$. Luminosity
  distances are calculated using the identified host galaxy redshift
  and the best fitting Planck 2013 cosmology ($\Omega_M=0.307$ and
  $H_0=67.8$ km s$^{-1}$ Mpc$^{-1}$), as implemented in the Astropy
  cosmology package.
\end{tablenotes}
\end{threeparttable}
\end{table*}

\section{Summary and Conclusions}
\label{sec:conc}

We calculate the radio emission from tidal disruption event jets
propagating through a range of plausible circumnuclear gas densities.
The latter are motivated by analytic estimates of the gas supply from
stellar winds based on our previous work in GSM15.  We simulate the
jet propagation using both 1D and 2D hydrodynamic simulations, which
we then post-process using a radiative transfer calculation to produce
synchrotron light curves. To isolate the effects of the density
profile and jet energy we employ a fixed two component jet model from
\citet{Mimica+2015}, which produces an acceptable fit to the observed
radio data of the on-axis jetted TDE SwJ1644.  Our conclusions are
summarized as follows.

\begin{enumerate}
\item The radio emission is most sensitive to the density at the jet
  deceleration radius, which is typically $r_{\rm dec} \sim 0.1-1$ pc
  (Fig.~\ref{fig:profiles}). We estimate the radial profile of nuclear
  gas densities expected from injection of stellar wind material for
  different star formation histories, and find that the gas density at
  10$^{18}$ cm lies in the range $n_{18} \sim 0.3 \Mbh[,7]^{-0.4} -
  1,300 \Mbh[,7]^{0.5}$ cm$^{-3}$, with $n_{18} \sim$ 10 cm$^{-3}$
  for star formation histories typical of TDE host galaxies (excluding
  a possible factor of $\sim$2 reduction from mass drop out from star
  formation).

\item The slope of the CNM gas density profile depends on the slope of
  the stellar density profile.  A TDE host galaxy likely possesses a
  cuspy stellar density profile inside of a few pc, with $\rho_\star
  \propto r^{-1.7}$. This translates into a gas density profile
  ranging from $n \propto r^{-0.7}$ on large scales to $n\propto
  r^{-1.5}$ on very small scales, well inside the stagnation radius,
  $r_{\rm s}$ and influence radius $r_{\rm inf}$. In general, we
  expect a density profile bracketed by $n\propto r^{-0.7}$ and
  $n\propto r^{-1.5}$ near the Sedov/deceleration radius. For
  simplicity we adopt a single power law $n\propto r^{-1}$ as our
  fiducial density profile.

\item We perform hydrodynamical simulations of our two component jet
  model for a range of plausible density profiles and normalizations
  $n_{18} = $ 2, 11, 60, 345, or 2000 cm$^{-3}$. We find bright radio
  emission at a few GHz across this entire range of densities. The
  peak luminosity is only weakly dependent on the chosen density
  profile for on-axis jets. For off-axis jets, the peak luminosity at
  1 GHz is insensitive to the CNM density profile and viewing angle
  for $n_{18}\geq 2$ cm$^{-3}$, although it will be a stronger
  function of density at higher frequencies.  While the peak radio
  flux is largely insensitive to the radial power-law slope for fixed
  $n_{18}$, a steeper profile $n\propto r^{-1.5}$ (e.g., as expected
  at radii $\ll r_{\rm s}, r_{\rm inf}$) alters the 2D dynamical
  evolution of the jet in a non-trivial way, resulting in a steeper
  post maximum decline of the radio light curve.

\item The time of the peak radio luminosity depends more sensitively
  on the density and can be as early as months, or as late as one
  decade, after the TDE. By comparing our calculated light curves with
  upper limits from a set of optical/UV and soft X-ray selected TDE,
  we show that most of these sources cannot have jets as powerful as
  SwJ1644.

\item In general, we only calculate the synchrotron radio
  emission from the forward shock, and neglect reverse shock
  emission. For high energy jets ($E\gsim 10^{53}$ erg), and
  frequencies $\lsim 30$ GHz, we find that the reverse shock has
  minimal impact on the total light curve. For low energy jets the
  reverse shock structure may be replaced by a series of recollimation
  shocks with a large emitting volume, which could contribute
  significantly to the total emission.

  Prompt radio follow-up, as well as regular monitoring, of future TDE
  flares would provide tighter constraints on the presence of jets.
  Radio afterglows can serve as calorimeters for off-axis jets
  launched by TDEs, and future observational efforts that capture the
  peak radio flux in thermally detected TDEs will add to the diversity
  of jet energies observed in TDE flares.  The broad range of energies
  (both detections and upper limits) already seen in TDE jets presents
  an interesting puzzle for theoretical models of jet launching.
\end{enumerate}

\section*{Acknowledgments}
We acknowledge helpful conversations with Jerry Ostriker, Luca Ciotti,
James Guillochon, and Yue Shen. BDM and AG acknowledge support from
the NSF (grant AST-1410950), NASA (grants NNX15AR47G, NNX15AU77G,
NNX16AB30G), the Alfred P.~Sloan Foundation, and the Research
Corporation for Science Advancement.  NS acknowledges support form the
NASA Einstein Postdoctoral Felloswhip Program (grant SAO
PF5-160145). DG acknowledges support from NASA through grant
NNX16AB32G issued through the Astrophysics Theory Program and support
from the Research Corporation for Science Advancement’s Scialog
program.  PM and MAA acknowledge financial support from the European
Research Council (ERC) through the Starting Independent Researcher
Grant CAMAP-259276, and the partial support of grants AYA2013-40979-P,
AYA2015-66899-C2-1-P and PROMETEO-II-2014-069. We thankfully
acknowledge the computer resources, technical expertise and assistance
provided by the Servei de Inform\`atica of the University of Valencia
and Columbia University's Yeti Computer Cluster. This research made
use of Astropy, a community-developed core Python package for
Astronomy (Astropy Collaboration, 2013).

\appendix
\section{Core Profile}
\label{app:core}
Fig.~\ref{fig:cores} compares the results of radio light curves from
jets propagating in core and cusp like gas density profiles
(Fig.~\ref{fig:profiles}).  We use the following analytic expression
to approximate the core galaxy CNM profile in Fig.~\ref{fig:profiles}
\begin{align}
\begin{cases}
n=n(r_s) k(x) & 0.4 \leq x\leq 2.0\\
n = 2.0 n(r_s) (x/0.4)^{-0.95} & x < 0.4\\
n = 0.75 n(r_s) (x/2.0)^{-0.26} & x>2,\\
\end{cases}
\label{eq:cores}
\end{align}
where
\begin{align}
  &x=r/r_s\\\nonumber
  &k(x)=\frac{45}{19} \frac{1}{x^{3/2}} \frac{1-x^{1.9}}{9-19
      x\frac{x^{0.9}-1}{x^{1.9}-1}}
\end{align}
To isolate the effects of the shape of the density profile, we
consider a core density profile with a stagnation radius $r_s=10^{18}$
cm and density normalization $n_{18}=2000$ cm$^{-3}$ which match those
of our high density cusp model.

\section{Peak Luminosities and times}
\label{app:analyt}
\citet{Leventis+2012} present analytic scaling relations for the
synchrotron flux of a spherical blast wave propagating through a
medium with a power law density profile, $n\propto r^{-k}$.  Here we
make use of their results to estimate the peak radio flux of the slow
(sheath) component of the jet.

During the late-time, Newtonian stage of the jet evolution,
synchrotron self absorption is important for frequencies below
\begin{align}
  \nu_{\rm sa}=&C_1(p, k) E_{54}^{\frac{10 p-k p -6 k}{2 (4+p) (5-k)}}
  n_{18}^{\frac{30 - 5 p}{2 (4 + p) (5 - k)}}
  \epsilon_e^{\frac{2 (p-1)}{4+p}} \epsilon_b^{\frac{p+2}{2 (4+p)}}\nonumber\\
  &t^{\frac{10 - 8 k - 15 p + 4 k p}{(4 + p) (5 - k)}},
\label{eq:nuSa} 
\end{align}
where $E = 10^{54}E_{54}$ erg is the blast wave energy and $C_1(p, k)$
is a normalization factor.  Equation~\eqref{eq:nuSa} is valid only if
self-absorption frequency is greater than the synchrotron peak
frequency,
\begin{equation}
\nu_m=C_2(p, k) E_{54}^{\frac{10-k}{2 (5-k)}} n_{18}^{-\frac{5}{2
    (5-k)}}  \epsilon_e^2  \epsilon_b^{1/2}  t^{\frac{4 k-15}{5-k}}.
\label{eq:num}
\end{equation}
The light curve will peak at the deceleration time (eq.~\ref{eq:tdec})
in case the emitting region is optically thin then. Otherwise, it will
occur after the deceleration time, when the self-absorption frequency
crosses through the observing band. The peak time for these two cases
is

\begin{align}
&t_{\rm p} \approx\nonumber\\
&\begin{dcases}
  0.5 \left(50 (3-k) \, E_{54}\right)^{1/(3-k)} \\\times\Gamma^{(2k - 8)/(3-k)}
  n_{18}^{-1/(3-k)} \,\,{\rm yr} & {\rm Opt.\, Thin}\\\\
  C_1(p, k)^{-\frac{(5 - k) (4 + p)}{10 - 8 k - 15 p + 4 k
      p}}E_{54}^{-\frac{-k p-6 k+10 p}{2 (4 k p-8 k-15 p+10)}}\\
  \times n_{18}^{-\frac{30-5 p}{2 (4 k p-8 k-15 p+10)}} \nu_{\rm
    obs}^{\frac{(5-k) (p+4)}{4 k p-8 k-15 p+10}}\\
  \times \epsilon_b^{-\frac{(5-k) (p+2)}{2 (4 k p-8 k-15 p+10)}}
  \epsilon_e^{-\frac{2 (5-k) (p-1)}{4 k p-8 k-15 p+10}} & {\rm Opt.\,
    Thick},
\end{dcases}
\label{eq:tpeakGen}
\end{align}
where $\Gamma$ is the initial jet Lorentz factor. 

The unabsorbed flux at the peak frequency is given by
\begin{align}
  F_{\nu_m} =  C_3(p, k) E_{54}^{\frac{8-3 k}{2 (5-k)}}
  n_{18}^{\frac{7}{2 (5-k)}} \epsilon_b^{1/2} t^{\frac{3-2 k}{5-k}}
\label{eq:Fnum}
\end{align}
Extrapolating to the observer frequency gives 
\begin{align}
  \nu_{\rm obs} F_{\rm p} (\nu_{\rm obs}) &= \nu_{\rm obs}   F_{\nu_m}
  \left(\frac{\nu_{\rm obs}}{\nu_m}\right)^{-(p-1)/2}.
  \label{eq:Fpeak1}
\end{align}
Combining equations~\eqref{eq:num}, ~\eqref{eq:tpeakGen}, ~\eqref{eq:Fnum},
and~\eqref{eq:Fpeak1}, we find
\begin{align}
  \nu_{\rm obs} F_{\rm p} (\nu_{\rm obs}) \propto
  \begin{dcases}
    E_{54}^{\frac{k (p+5)-12}{4 (k-3)}} n_{18}^{-\frac{3 (p+1)}{4
        (k-3)}} \nu_{\rm obs}^{\frac{3-p}{2}}
    \epsilon_b^{\frac{p+1}{4}} \epsilon_e^{p-1} & {\rm Opt.\, Thin}\\\\
    E_{54}^{\frac{k(-(p-2))-10 p+3}{4 k (p-2)-15 p+10}} \\ \times
    n_{18}^{\frac{11 (p-2)}{4 k (p-2)-15 p+10}} \nu_{\rm
      obs}^{\frac{14 k (p-2)-47 p+57}{4 k (p-2)-15 p+10}} \\ \times
    \epsilon_b^{\frac{k (-(p-2))+p-8}{4 k(p-2)-15 p+10}}
    \epsilon_e^{-\frac{11 (p-1)}{4 k (p-2)-15p+10}} & {\rm Opt.\,
    Thick}
  \end{dcases}
  \label{eq:peakLumGen}
\end{align}

After peak, we expect that the flux scales as 

\begin{equation}
F_{\nu}\propto t^{\frac{21-8k-15p+4kp}{10-2k}}.
\label{eq:tslope}
\end{equation}

\section{Reverse shock}
\label{sec:reverse}
Here we estimate the fraction of the kinetic energy of the jet that is
dissipated by the reverse shock, as opposed to the forward shock whose
contribution is the focus of this paper.  From continuity, the
comoving density of a relativistic jet is given by
(e.g.~\citealt{Uhm&Beloborodov2007})
 \begin{align}
   n_{\rm j} =  \frac{L_{\rm j, iso}}{4 \pi r^{2}\Gamma_{\rm
       j}^{2}c^{3}m_p(1 + r \dot{\Gamma}/c\Gamma^{3})}
   \approx  \frac{L_{\rm j, iso}}{4 \pi r^{2}\Gamma^{2}c^{3}m_p},
\end{align}
where $L_{\rm j, iso}$ is the isotropic equivalent luminosity.  The
second term in the denominator can be neglected if the jet Lorentz
factor changes slowly ($\dot{\Gamma}_{\rm j} \ll c\Gamma^{3}/r$), a
condition which is satisfied at radii $r < r_{\rm dec}$ if $\Gamma$
changes slowly on a timescale $\gtrsim t_{\rm 0}$, where $t_{\rm 0}$
is the jet duration.

The common Lorentz factor of the shocked CNM and the shocked jet can
be estimated using the relativistic shock jump condition and pressure
equality between the forward and reverse shocks.  In the
ultra-relativistic limit this gives,
\begin{equation}
\Gamma_{\rm sh} \underset{\Gamma_{\rm sh} \gg 1}= \Gamma\left[1 +
  2\Gamma f^{-1/2}\right]^{-1/2},
\label{eq:gammaShSim}
\end{equation}
where
\begin{equation}
  f\approx 40\,  L_{\rm j,48} n_{18}^{-1} \Gamma_{10}^{-2} \, \left(\frac{r}{10^{18} {\rm
        cm}}\right)^{-1} 
\end{equation}
is the ratio of the density of the jet to that of the
CNM. Equation~\eqref{eq:gammaShSim} is inaccurate for mildly relativistic or
non-relativistic flows, in which case we apply the more general
expression for $\Gamma_{\rm sh}$ given by \citet{Beloborodov&Uhm2006}
(their eq.~3, see also \citealt{Mimica&Aloy2010})
\begin{equation}
\frac{\Gamma_{\rm sh}^2-1}{\Gamma_{43}^2-1} f^{-1}=1 ,
\label{eq:gamma43}
\end{equation}
where
\begin{equation}
  \Gamma_{43}=\Gamma \Gamma_{\rm sh} \left(1-\beta_{\rm sh} \beta_j\right),
\label{eq:gammaShGen}
\end{equation}
is the Lorentz of shocked jet in the frame of the unshocked
jet. Combining equations~\eqref{eq:gamma43}
and~\eqref{eq:gammaShGen}, we obtain 

\begin{align}
&\Gamma_{\rm sh}(f)=\sqrt{\frac{f \left(\Gamma ^2 (f-3)-2 \left(\Gamma ^2-1\right) \Gamma
      \sqrt{f}+1\right)+1}{(f+1)^2-4 \Gamma ^2 f}}\nonumber\\
&\Gamma_{43}(f) = \sqrt{\frac{4 \Gamma  f^{3/2}+f^2+\Gamma ^4 f+4 \Gamma ^3 \sqrt{f}+2 \Gamma ^2 (2 f+1)+f-1}{\left(2 \Gamma  \sqrt{f}+f+1\right)^2}}
\label{eq:gammaShGen2}
\end{align}
In the lab frame the reverse shock moves with a velocity
\begin{equation}
\beta_{\rm rs}=\frac{\beta_{\rm sh}(f)-\beta_{43}(f)/3}{1-\beta_{\rm
    sh}(f) \beta_{43}(f)/3}.
\label{eq:betars}
\end{equation} 
Equations~\eqref{eq:gammaShGen2} and ~\eqref{eq:betars} can be used to
determine the radius of the shocks when the reverse shock crosses the
trailing edge of the jet and the value of $\Gamma_{\rm sh, rs}$ at this
time.  This involves numerically integrating $\beta_{\rm
  rs}/\beta_{j}=d r_{\rm rs}/d r_{\rm ej}$, where $r_{\rm \rs}$ is the
position of the reverse and $r_{\rm ej}$ is the position of the back
of the jet. The latter allows us to calculate what fraction of the
initial kinetic energy of the jet is dissipated at the reverse shock,
instead of being transferred to the shocked external medium via the
forward shock. This is approximately given by

\begin{equation}
f_{\rm ke} \approx\frac{\Gamma-\Gamma_{\rm sh, rs}}{\Gamma-1}
\end{equation}

\clearpage
  \footnotesize{
    \bibliographystyle{mnras}
    \bibliography{master}
  }

\end{document}